\documentclass[%
 aip,
 amsmath,amssymb,
 reprint,%
]{revtex4-1}

\usepackage[utf8]{inputenc}
\usepackage{graphicx}
\usepackage{dcolumn}
\usepackage{bm}
\usepackage{hyperref}
\usepackage[mathlines]{lineno}
\usepackage{amsmath} 
\usepackage{siunitx}   
\usepackage{multirow}  
\usepackage{textcomp}  

\relax 

\begin{document}

\title{Taiwan Axion Search Experiment with Haloscope: Designs and operations}

\author{Hsin~Chang}\affiliation{Department of Physics, National Central University, Taoyuan City 320317, Taiwan}
\author{Jing-Yang~Chang}\affiliation{Department of Physics, National Central University, Taoyuan City 320317, Taiwan} 
\author{Yi-Chieh~Chang}\affiliation{National Synchrotron Radiation Research Center, Hsinchu 300092, Taiwan} 
\author{Yu-Han~Chang}\affiliation{Department of Physics, National Chung Hsing University, Taichung City 402202, Taiwan}
\author{Yuan-Hann~Chang}\affiliation{Institute of Physics, Academia Sinica, Taipei City 115201, Taiwan}
\affiliation{Center for High Energy and High Field Physics, National Central University, Taoyuan City 320317, Taiwan}
\author{Chien-Han~Chen}\affiliation{Institute of Physics, Academia Sinica, Taipei City 115201, Taiwan} 
\author{Ching-Fang~Chen}\affiliation{Department of Physics, National Central University, Taoyuan City 320317, Taiwan}
\author{Kuan-Yu~Chen}\affiliation{Department of Physics, National Central University, Taoyuan City 320317, Taiwan} 
\author{Yung-Fu~Chen}\email[Correspondence to: ]{yfuchen@ncu.edu.tw}\affiliation{Department of Physics, National Central University, Taoyuan City 320317, Taiwan} 
\author{Wei-Yuan~Chiang}\affiliation{National Synchrotron Radiation Research Center, Hsinchu 300092, Taiwan}
\author{Wei-Chen~Chien}\affiliation{Department of Physics, National Chung Hsing University, Taichung City 402202, Taiwan}
\author{Hien~Thi~Doan}\affiliation{Institute of Physics, Academia Sinica, Taipei City 115201, Taiwan} 
\author{Wei-Cheng~Hung}\affiliation{Department of Physics, National Central University, Taoyuan City 320317, Taiwan}\affiliation{Institute of Physics, Academia Sinica, Taipei City 115201, Taiwan} 
\author{Watson~Kuo}\affiliation{Department of Physics, National Chung Hsing University, Taichung City 402202, Taiwan} 
\author{Shou-Bai~Lai}\affiliation{Department of Physics, National Central University, Taoyuan City 320317, Taiwan} 
\author{Han-Wen~Liu}\affiliation{Department of Physics, National Central University, Taoyuan City 320317, Taiwan} 
\author{Min-Wei~OuYang}\affiliation{Department of Physics, National Central University, Taoyuan City 320317, Taiwan}
\author{Ping-I~Wu}\affiliation{Department of Physics, National Central University, Taoyuan City 320317, Taiwan} 
\author{Shin-Shan~Yu}\affiliation{Department of Physics, National Central University, Taoyuan City 320317, Taiwan}
\affiliation{Center for High Energy and High Field Physics, National Central University, Taoyuan City 320317, Taiwan}


\date{\today}

\begin{abstract}  

We report on a holoscope axion search experiment near $19.6\ \si{\micro \eV}$ from the Taiwan Axion Search Experiment with Haloscope collaboration. The experiment is carried out via a frequency-tunable cavity detector with a volume $V = 0.234\ {\rm liter}$ in a magnetic field $B_0 = 8\ {\rm T}$. With a signal receiver that has a system noise temperature $T_{\rm sys} \cong 2.2\ {\rm K}$ and experiment time about 1 month, the search excludes values of the axion-photon coupling constant $g_{\rm a\gamma\gamma} \gtrsim 8.1 \times 10^{-14} \ {\rm GeV}^{-1}$, a factor of 11 above the Kim-Shifman-Vainshtein-Zakharov benchmark model, at the 95\% confidence level in the mass range of $19.4687-19.8436\ \si{\micro \eV}$. We present the experimental setup and procedures to accomplish this search.

\end{abstract}

\date{\today}

\maketitle

\section{Introduction}

\begin{figure}[tb]
    \centering
    \includegraphics[width=0.41\textwidth]{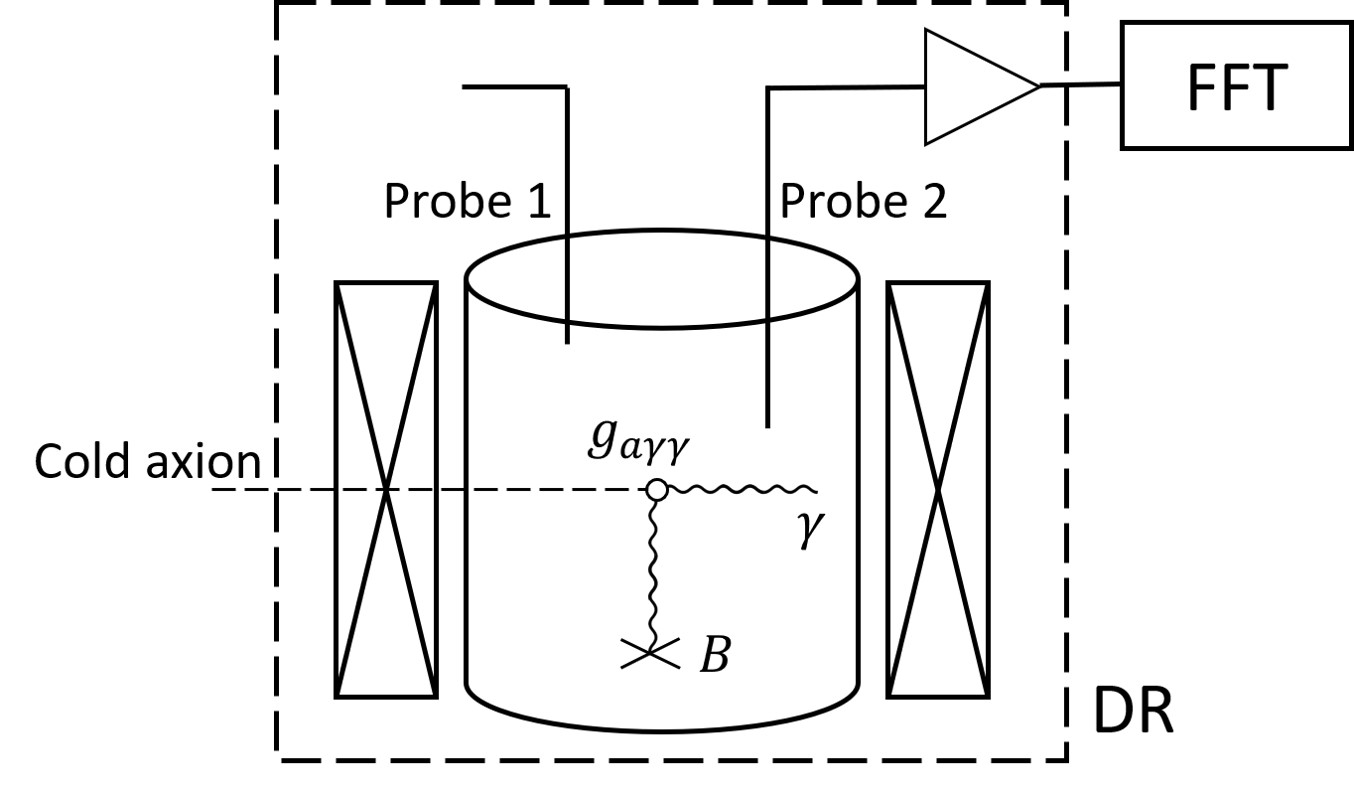}
    \caption{Schematic diagram of axion haloscope detection. An incoming axion couples to the provided magnetic field $B$ and converts to a photon. The photon is captured by a MW cavity and detected through the readout probe 2. An additional weak coupling probe 1 is used for examining the cavity characteristics and verifying the detection system. A DR hosts the cavity at a millikelvin environment and the magnet.}
    \label{fig:haloscope}
\end{figure}

The axion is a hypothetical particle predicted as a consequence of a solution, proposed by Peccei and Quinn, to the strong CP problem in quantum chromodynamics (QCD)~\cite{peccei1977cp,peccei1977constraints,weinberg1978new,wilczek1978problem}, and is also considered as a compelling candidate for the cold dark matter~\cite{duffy2009axions}. The coupling of the axion to the standard-model (SM) particles is model-dependent, but generally very weak~\cite{kim1979weak,shifman1980can,dine1981simple,zhitnitskij1980possible}. Recent astrophysical and cosmological considerations suggest a mass range of $\mathcal{O}(1-100)\ \si{\micro \eV}$ for the axion~\cite{QCDCalI, QCDCalII, QCDCalIII, QCDCalIV, QCDCalV, QCDCalVI, QCDCalVII, QCDCalVIII, QCDCalIX, QCDCalX, QCDCalXI, QCDCalXII, QCDCalXIII, bradley2003microwave}. The existence of the axion is yet not confirmed experimentally. Pierre Sikivie proposed a microwave (MW) cavity experiment (called haloscope) to detect the axion dark matter in a laboratory~\cite{sikivie1983experimental,sikivie1985detection}. Figure \ref{fig:haloscope} shows a schematic of the MW cavity axion haloscope detection. Via a two-photon coupling process, an axion of mass $m_{\rm a}$ in a static magnetic field $B$ can convert to an equal-energy photon with a frequency $f_{\rm a} \cong m_{\rm a}c^2/h$, where $c$ is the speed of light and $h$ is the Planck constant. The converted photons can be accumulated in a cavity with the resonant frequency $f_{\rm c}$ matched with $f_{\rm a}$ and subsequently be detected by a signal receiver through the probe 2 with adequate coupling to the cavity. In addition, the weak coupling probe 1 to the cavity is introduced for testing purposes. Although the haloscope is a narrowband detection and requires tuning the cavity frequency, to date it is the only detection scheme sensitive to the QCD prediction limits in the preferential axion mass range. 

The expected axion-photon conversion signal power from a frequency-matched cavity to the readout probe 2 is given by~\cite{sikivie2021invisible}
\begin{equation}
\label{eq:axion_signal_power}
    P_{\rm a} = \eta g_{\rm a\gamma\gamma}^2 \left( \frac{\rho_{\rm a}}{m_{\rm a}} \right) B_0^2 V C_n Q.
\end{equation}
Here, $g_{\rm a\gamma\gamma}=(g_{\rm \gamma}\alpha/\pi\Lambda^2)m_{\rm a}$ is the axion-photon coupling constant, where $\alpha$ is the fine-structure constant, $\Lambda$ = 78 MeV is a scale parameter that can be derived from the mass and the decay constant of the pion and the ratio of the up to down quark masses, and $g_{\rm \gamma}$ is the model-dependent parameter. The values of $g_{\rm \gamma}$ are -0.97 and 0.36 in the KSVZ (Kim-Shifman-Vainshtein-Zakharov)~\cite{kim1979weak,shifman1980can} and the DFSZ (Dine-Fischler-Srednicki-Zhitnitskii)~\cite{dine1981simple,zhitnitskij1980possible} benchmark models, respectively. $\rho_{\rm a} = 0.45\ {\rm GeV/cm^3}$ is the local dark matter density~\cite{read2014local,particle2020review}. $B_0$ is the nominal magnetic field strength. $V$ is the cavity volume. $C_n < 1$ is the cavity-mode-$n$-dependent form factor, which will be described in Sec. \ref{subsec:Cavity design} and Eq. \ref{eq:cavity form factor}. $Q = (1/Q_0 + 1/Q_1 + 1/Q_2)^{-1} = 2\pi f_{\rm c}/\kappa$ is the cavity overall quality factor, where $Q_0 = 2\pi f_{\rm c}/\kappa_0$ is related to the cavity intrinsic loss rate $\kappa_0$, $Q_{1,2} = 2\pi f_{\rm c}/\kappa_{1,2}$ are related to the loss rate through the probe 1, $\kappa_1$, and the emission rate to the signal receiver via the probe 2, $\kappa_2$, respectively. We also define the total loss rate $\kappa = \kappa_0 + \kappa_1 + \kappa_2$, and the coupling coefficients $\beta_{1,2} = \kappa_{1,2}/\kappa_0 = Q_0/Q_{1,2}$. Finally, $\eta = \kappa_2/\kappa = \beta_2/(1+\beta_1+\beta_2)$ is the fraction of the converted power transmitting to the readout probe 2. 

Based on Eq. \ref{eq:axion_signal_power}, the expected signal power for this experiment is of the order of $10^{-24}\ {\rm W}$ or smaller. A low-noise, linear amplification chain providing substantial amplification is essential to the signal receiver of a haloscope. According to the Dicke radiometer equation~\cite{dicke1946measurement}, the fluctuation in the average noise power for any signal receiver is given by 
\begin{equation}
\label{eq:Dicke_eq}
    \sigma_{\rm n} = k_{\rm B} T_{\rm sys} \sqrt{\frac{\Delta f}{t}}, 
\end{equation}
where $k_{\rm B}$ is the Boltzmann constant, $T_{\rm sys}$ is the system noise temperature, $\Delta f$ is the bandwidth, and $t$ is the integration time. $T_{\rm sys} = \Tilde{T}_{\rm c} + T_{\rm a}$ includes contributions from the background noise of the cavity and the added noise of the receiver electronics. $\Tilde{T}_{\rm c} = \frac{hf}{k_{\rm B}} \left( \frac{1}{e^{hf/k_{\rm B}T_{\rm c}} - 1} +\frac{1}{2} \right)$ is the effective noise temperature of the cavity and consists of the thermal and the vacuum noises, where $T_{\rm c}$ is the physical temperature of the cavity, and $T_{\rm a}$ is the added noise temperature of the receiver electronics and is mainly determined by the noise performance of the first stage amplifier. High magnetic field environment, large cavity volume, efficient cavity mode, high cavity quality, low system noise, and long integration time, will therefore maximize the detection signal-to-noise ratio (SNR) $P_{\rm a}/\sigma_{\rm n}$. Furthermore, the cavity resonance frequency tuning step $\Delta f_{\rm s}$ is optimized at an half of the cavity bandwidth $\Delta f_{\rm c}/2 = \kappa/4\pi$~\cite{al2017design}. The search scan rate $\Delta f_{\rm s}/t$ with a fixed detection limit is the most important figure of merit of a cavity haloscope. The choice of the coupling coefficient $\beta_2 = 2$ maximizes the search scan rate in the limit of $\beta_1 \ll 1$ ~\cite{al2017design}. 

Several programs have actively carried out axion haloscope searches, and a few have reached or approached the QCD limit. The most significant efforts were from the ADMX, which excluded the KSVZ benchmark model within the mass range of $1.9-4.2\ \si{\micro \eV}$~\cite{asztalos2010squid,du2018search,braine2020extended,bartram2021search}. The CAPP have worked on the range below $15\ \si{\micro \eV}$~\cite{lee2020axion,kwon2021first}, and excluded the KSVZ model within $10.7126-10.7186\ \si{\micro \eV}$~\cite{kwon2021first}. For higher mass ranges, the search becomes less efficient due to the small cavity volume at the corresponding resonant frequency.

The phase-insensitive linear signal receiver, ultimately with the noise performance subject to quantum fluctuations, is chosen in typical haloscopes~\cite{lamoreaux2013analysis}. The lately developed Josephson parametric amplifiers (JPAs)~\cite{castellanos2008amplification,yamamoto2008flux} have achieved quantum-limited added noise in the $0.5-10\ {\rm GHz}$ frequency range. By integrating JPAs to the cavity haloscope signal receiver, several programs, including the HAYSTAC~\cite{brubaker2017first}, the QUAX~\cite{alesini2021search}, and the ADMX~\cite{bartram2021dark}, have begun to look for axions above $15\ \si{\micro \eV}$. A work from the HAYSTAC has achieved a search of excluding axion-photon coupling near the QCD limit in a $0.3\ \si{\micro \eV}$ mass window around $17\ \si{\micro \eV}$~\cite{backes2021quantum}.

The TASEH (Taiwan Axion Search Experiment with Haloscope) is a haloscope axion dark matter search experiment sited in Taiwan. Based on our projected developments of the cavity detectors under the constraint of our future magnet availability, we target to search axion in the $10-25\ \si{\micro \eV}$ mass range, roughly corresponding to the $2.5-6\ {\rm GHz}$ MW frequency band, up to the QCD axion-photon coupling limit. In this frequency range, the signal receiver consisting of a linear amplifier detector is still the optimal choice~\cite{lamoreaux2013analysis}. The ultimate plan of TASEH is to integrate a quantum-limited JPA to the signal receiver and to develop a tunable large-volume cavity detector to boost the haloscope axion search in this mass range.

TASEH has conducted the first successful axion detection run in October-November 2021, which is termed as the CD102 run, where CD stands for "cool down". This run excludes values of the axion-photon coupling constant $g_{\rm a\gamma\gamma} \gtrsim 8.1 \times 10^{-14} \ {\rm GeV}^{-1}$ at the 95\% confidence level in the mass range of $19.4687-19.8436\ \si{\micro \eV}$. The exclusion limit is about a factor of 11 above the KSVZ model. In this paper, we introduce the experimental setup and the operation procedures to accomplish this detection run, and present the experiment results. We refer to our parallel paper~\cite{TASEH2022analysis} for the detailed analysis procedures. 
\section{Experimental setup}

This section describes our axion haloscope instrumentation setup. Figure \ref{fig:haloscope} shows the schematic of our haloscope. A frequency-tunable and high-quality MW cavity accumulates photons converted from axions in a strong magnetic field $B$ produced by a superconducting solenoid magnet. A cryogen-free dilution refrigerator (DR) hosts the cavity at a millikelvin environment and the magnet. The signal from the cavity is directed to a low-noise amplification chain, with a high-electron-mobility-transistor (HEMT) amplifier in the first stage. The amplified signal is downconverted and sampled by a vector signal analyzer to retrieve the total spectrum. The following subsections \ref{subsec:DR&Magnet} - \ref{subsec:Microwave system} report the important components of the experimental setup respectively.

\subsection{Dilution refrigerator and magnet}\label{subsec:DR&Magnet}

The axion haloscope detection experiment requires a cryostat for several purposes. The cavity detector needs to sit in a cryogenic environment in order to reduce its thermal noise down to the vacuum fluctuation limit. As the cavity temperature $T_{\rm c}$ is considerably lower than the half photon energy of interest, i.e. $T_{\rm c} \ll hf_{\rm c}/2k_{\rm B} \equiv T_{\rm q}/2$, $\Tilde{T}_{\rm c}$, which is proportional to the emitted noise power of the cavity, saturates at the level $T_{\rm q}/2$. For a 5-GHz cavity, $T_{\rm q}/2$ is about 0.12 K. The operation of a state-of-the-art low-noise HEMT amplifier is typically at 4 K. Future integration of quantum-limited JPAs into the signal receiver also requires millikevin background. Besides, the large superconducting magnet must be cooled down to about 4 K for normal operation. A DR cryostat therefore is essential for conducting our axion detection experiment.

\begin{figure}[tb]
    \centering
    \includegraphics[width=0.48\textwidth]{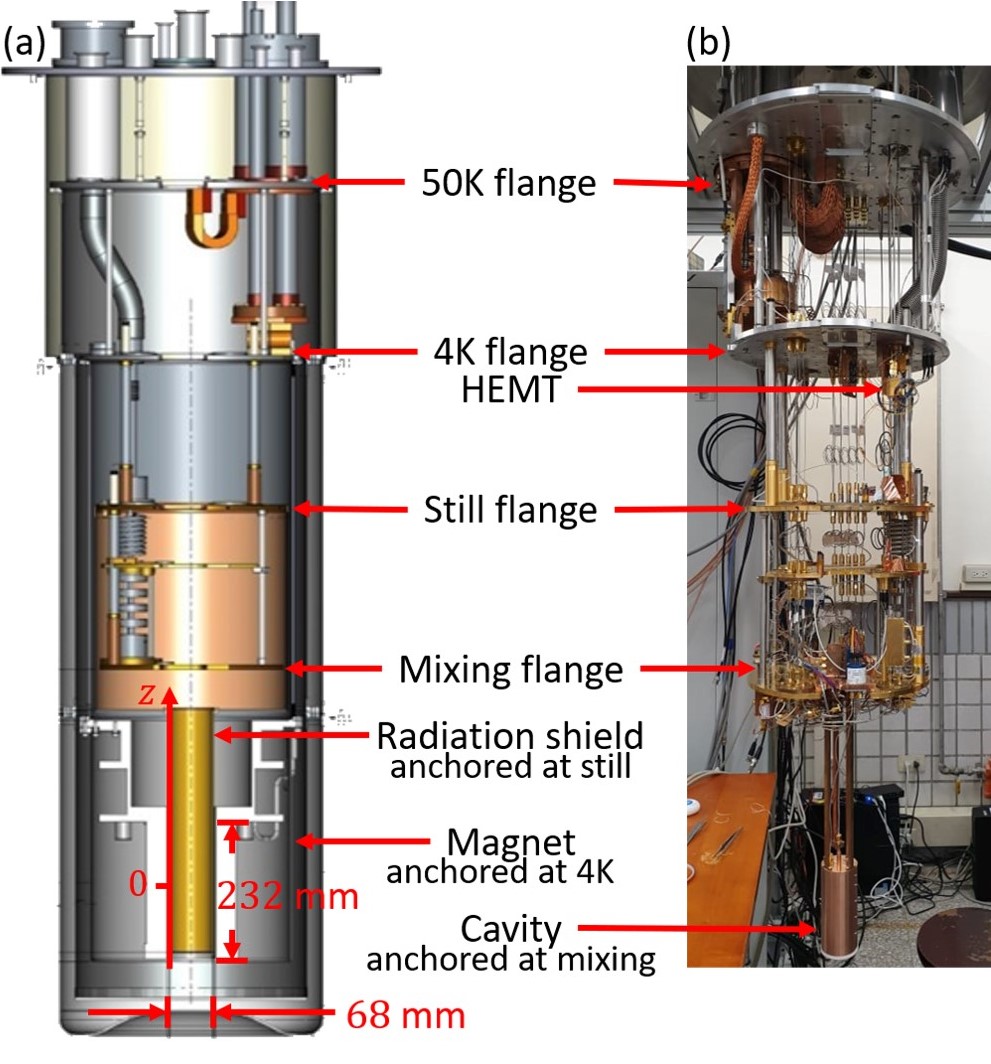}
    \caption{TASEH axion detection experimental setup. (a) Schematic diagram of cryogen-free DR system. (b) Overall view picture of experimental setup.}
    \label{fig:DR_magnet}
\end{figure}

Figure \ref{fig:DR_magnet} shows a schematic of our cryogen-free DR system~\cite{BlueFors2014manual}, built by Bluefors in 2012 (model: BF-LD250), and the corresponding picture of the experimental setup. The system is equipped with a Cryomech PT410-RM pulse tube cryocooler (PTC), which provides 35 W cooling power at 45 K in the first stage and 1.0 W at 4 K in the second stage. The DR 50K and 4K flanges are anchored to the first and the second stages of PTC, respectively. The system requires approximately 18/93 liters of He3/He4 mixture gas in the circulation path. With our 8T superconducting magnet attached, the initial DR cool down time to the base temperature is roughly 48 hours. In the normal circulation condition, the still, exchange, and mixing flanges provide cooling powers of 8 mW at 850 mK, $100\ \si{\micro\watt}$ at 140 mK, and $10\ \si{\micro\watt}$ at 20 mK, respectively. With the axion experiment heat loads, the system base temperature at the mixing flange $T_{\rm mx}$ is approximately 27 mK. Besides precooling the mixture, the 4K flange cooling power is particularly important to take away the radiation heat load absorbed by the superconducting magnet surface area, the HEMT bias heating, and the measurement wiring heat load. The mixing flange cooling power is to remove the measurement wiring heat load and the heat from the motor and MW switch operations. 

The DR equips 7 thermometers in various spots to monitor the system operation; see Table \ref{table:tablethermometer} in Appendix \ref{sec:Table} for details. 48 channels of dc wiring from the room temperature vacuum flange to the mixing flange are available for supporting cryogenic component operations in the mixing flange, such as MW switches and piezo motors; see Table \ref{table:tabledcwiring} for details.

The superconducting solenoid magnet made of NbTi wire is manufactured by American Magnetics Inc. The magnet is thermally anchored at the DR 4K flange through an interface tube, made of high purity aluminum, to ensure that it operates at 4 K. The magnet bore diameter and length are 76 and 240 mm, respectively. A gold-plated radiation shield anchored at the still flange fits into the bore to prevent the cavity in the bore from seeing the magnet thermal radiation. The radiation shield inner diameter and length are 68 and 232 mm, respectively, which limit the sizes of the cavity fitting into the magnet, and therefore set the lower bound of the cavity resonance frequency.

\begin{figure}[tb]
    \centering
    \includegraphics[width=0.3\textwidth]{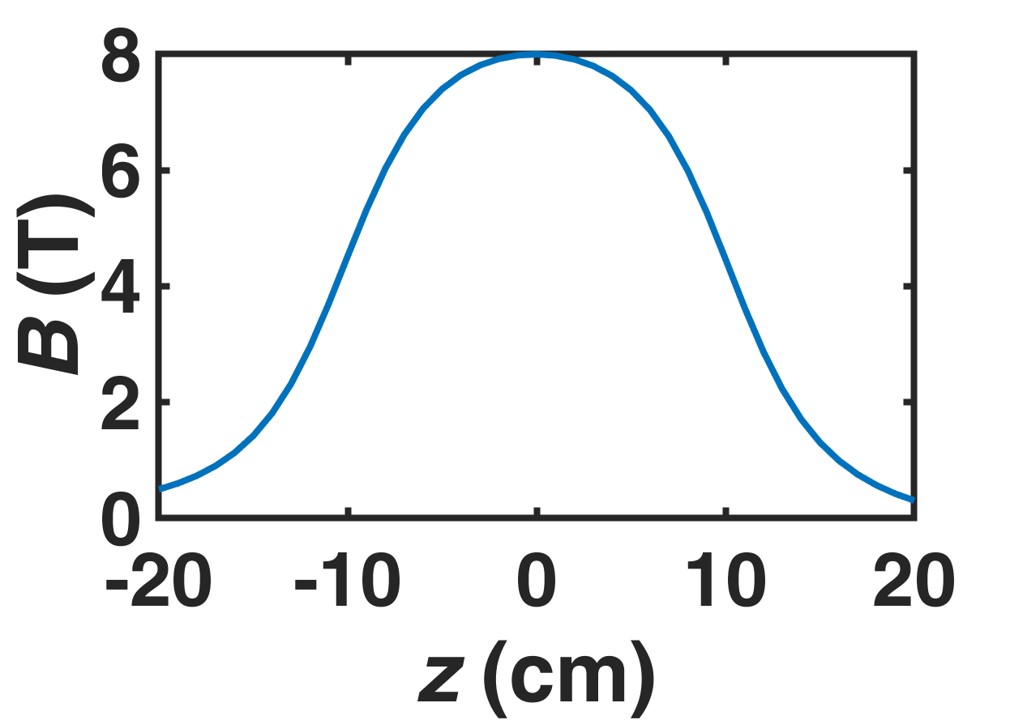}
    \caption{Magnetic field distribution along axial direction.}
    \label{fig:magnet}
\end{figure}

The magnet achieves the nominal central field $B_0=$ 8 T when charged at 86.21 A. Figure \ref{fig:magnet} shows the field distribution in the bore along the axial direction (z-direction)~\cite{BlueFors2014manual}. $z = 0$ marks the magnet center. The field within $\pm$60 mm is greater than 7 T when fully charged. The typical ramp rate of the magnet is 100 mT/min. With this rate it takes 80 minutes to charge to, or discharge from, the full field. The magnet is equipped with a persistent switch to support the persistent mode operation. The magnet center is 356 mm below the mixing flange. A compensation coil helps to reduce the field near the mixing flange below 40 mT when the magnet is fully charged. The compensated region is within 51 mm below the mixing flange. The feature is useful for the operation of ferromagnetic MW circulators, and could be particularly important when integrating quantum-limited JPAs into the signal receiver in the future. 

A magnet quench occurs whenever the helium compressor of PTC stops working, typically due to an electrical power outage or a cooling water failure. With the 12-Henry magnet inductance, 45 kJ is stored in the magnetic field when the magnet is fully charged. The stored energy dissipates into the resistive NbTi wire when a quench happens. This causes the entire system with the experiment apparatus (except the 50K flange) to warm up to above 30 K in a few minutes. During the CD102 run, one quench occurred due to a cooling water failure. There was no damage in the experimental setup in the course of this quench event although all the mixture evaporated and went back to the storage tank. It takes about 7 hours to cool down the system back to the base temperature and additional 1.5 hours to resume the 8T magnetic field. To prevent a magnet quench, the cooling water temperature is monitored and an alarm will be triggered if the temperature is abnormally high. 

\subsection{Microwave cavity}\label{subsec:MW cavity}

The axion detection figure of merit largely depends on the cavity performance. Enlarging the cavity $V$, $C_n$, and $Q_0$ can enhance the axion signal. The design targets the cavity frequency at around 5 GHz, and maximizes $V$ and $C_n$ with the constraint of the magnetic bore dimensions. High purity copper and surface treatments are used to fabricate the cavity body and most of the components to raise $Q_0$. The Ansys HFSS MW simulation guides the design, and the scattering parameter (S-parameter) measurement determines the cavity performance.

\subsubsection{Design}\label{subsec:Cavity design}

The form factor $C_n$ is the normalized overlap of the electric field $\textbf E_n$, for a particular cavity resonant mode $n$, with the external magnetic field $\textbf B$:
\begin{equation}
    \label{eq:cavity form factor}
    C_n = \frac {(\int \textbf B \cdot \textbf E_n dV)^2} {B^2_0 V \int |\textbf E_n^2| dV}.
\end{equation}
The field $\textbf B$ inside the magnet bore points mostly along the z-direction. The cylindrical cavity utilizes the magnet bore volume most efficiently. The most favorable mode is TM$_{010}$, whose electric field is parallel to the magnetic field in the magnet bore without any node. 

\begin{figure}[tb]
    \centering
    \includegraphics[width=0.48\textwidth]{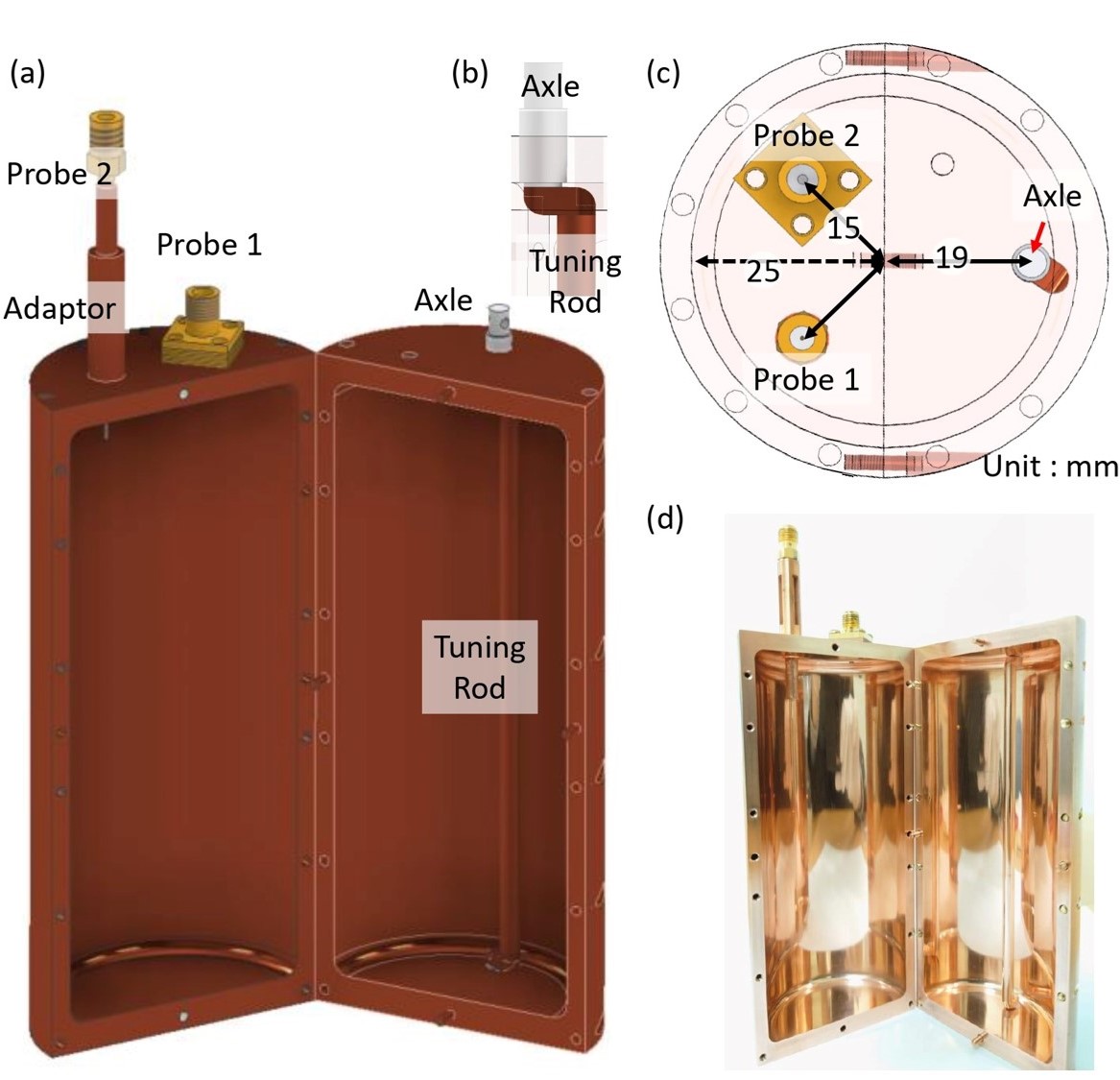}
    \caption{Cavity design. (a) Cavity layout. (b) Zoom-in view of frequency tuning rod near top axle. (c) Top view. (d) Photo of experimental object.}
    \label{fig:cavity}
\end{figure}

Figure \ref{fig:cavity}(a) shows the schematic layout of the frequency-tunable copper cavity. The cavity has an off-axis rod inside for frequency tuning and two probes introduced from the top. The frequency tuning rod is made of copper. Two  axles made of teflon are fixed at the rod ends and fit to the cavity top and bottom walls; see Fig. \ref{fig:cavity}(b) for a zoom-in view near the top teflon axle. The teflon axles, which are insulated from the cavity wall, avoid the rod acting like an antenna, and allow the rod to rotate along the off-axis freely. The top axle is connected to a rotational motor to achieve frequency tuning due to the reduction of the effective cavity diameter. The probe 1 has weak fixed coupling for examining the cavity characteristics and testing the signal receiver. The probe 2 has tunable coupling for optimizing the axion signal detection. The tuning mechanism is achieved via a sliding probe connected to a linear motor to adjust its insertion depth. For the TM$_{010}$ mode, the current flows in the cavity inner wall along the longitudinal direction. The cavity is split into two halves along the longitudinal direction to reduce the loss from the seam~\cite{choi2021capp}.

Considering the mechanical strength of the cavity, we set the cavity wall thickness at 4 mm. The thickness, together with the inner diameter of the radiation shield, limits the cavity inner diameter $D$ = 50 mm. Without a tuning rod inside, the expected $f_{\rm c}$ = 4.56 GHz for the TM$_{010}$ mode is right at our target range. The cavity length $L$ = 120 mm is chosen to fully exploit the available high magnet field region and to avoid any TE mode coming close to the TM$_{010}$ mode. Figure \ref{fig:cavity}(c) shows the top view of the cavity layout. The two probes are 15 mm away from the center and 90\textdegree~away from each other. The frequency tuning rod has 4 mm in diameter and 114 mm in length. The rotation axis of the tuning rod is 19 mm away from the center and 135\textdegree~away from the two probes. The off-axis length of the rod is 3.5 mm and it can rotate the full 360\textdegree. A 1-mm gap is left between the transverse part of the rod end and the top/bottom wall. With the frequency tuning rod inside, the cavity volume $V$ = 0.234 liter.

\subsubsection{Simulation}

\begin{figure}[tb]
    \centering
    \includegraphics[width=0.48\textwidth]{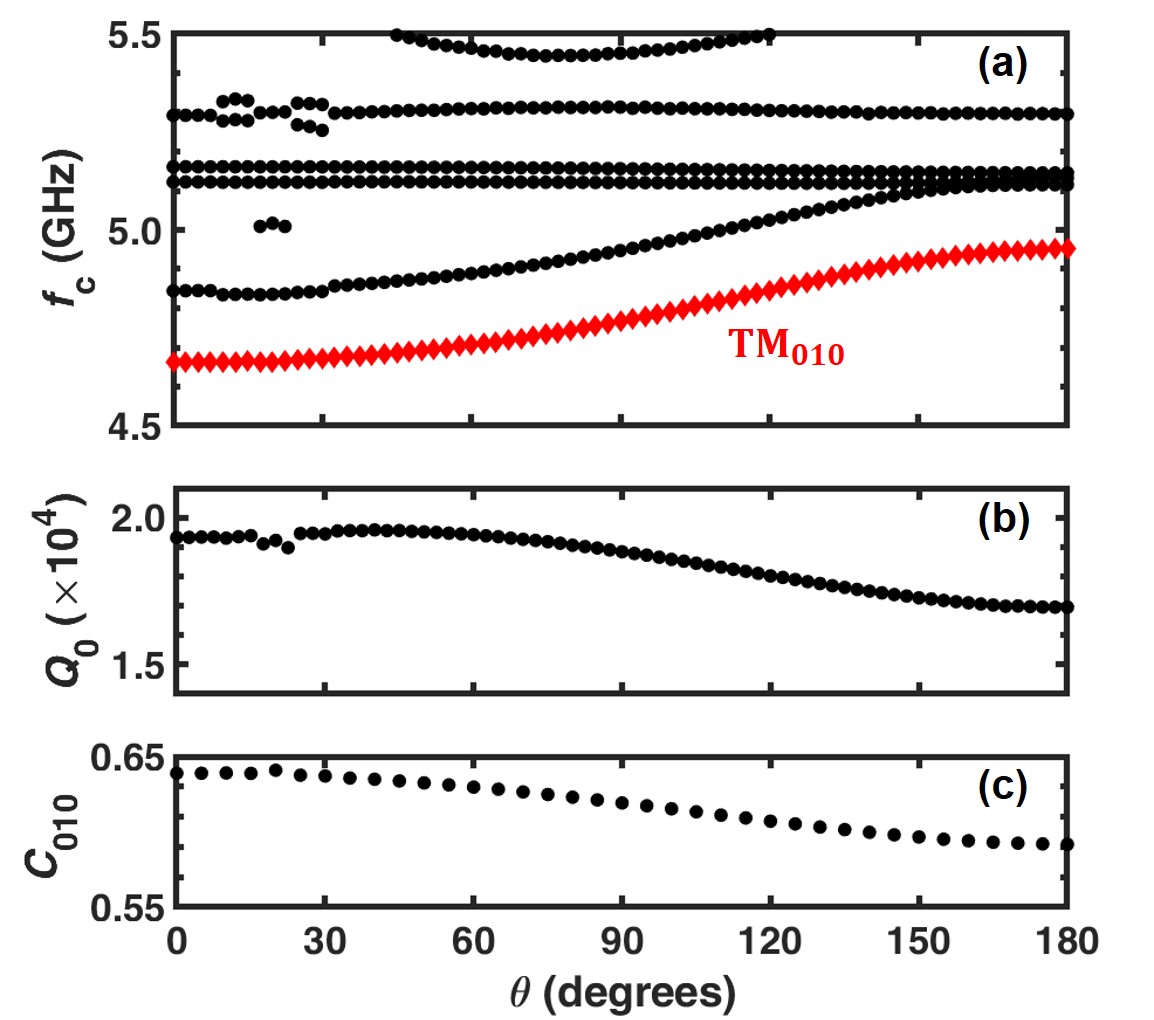}
    \caption{Ansys HFSS simulation of the TASEH frequency-tunable cavity. (a) Mode map; frequencies vs. $\theta$. The red diamonds mark the TM$_{010}$ mode. (b) Intrinsic quality factor $Q_0$ of TM$_{010}$ mode vs. $\theta$. (c) Form factor $C$ of TM$_{010}$ mode vs. $\theta$. The variation in the search frequency range of the CD102 run is within $0.62-0.63$.}
    \label{fig:SimModemap}
\end{figure}

We build the model of the cavity by the Ansys HFSS, a finite element based MW simulation software, to simulate the room temperature characteristics of the cavity. Figure \ref{fig:SimModemap} shows the simulation results, including the resonance frequencies of the TM$_{010}$ mode and the nearby modes, as well as the TM$_{010}$ mode form factor $C$ and intrinsic quality factor $Q_0$, as functions of the tuning rod angle $\theta$. The TM$_{010}$-like mode is labeled by the red diamonds in Fig. \ref{fig:SimModemap}(a), determined by the electric field distribution of this mode. The mode frequency $f_{\rm c}$ increases from 4.662 to 4.951 GHz as $\theta$ moves from 0\textdegree~to 180\textdegree, with an approximate slope of 2 MHz/degree in the 50\textdegree$-$100\textdegree~$\theta$ range. This mode manifests no mode crossing in the entire $\theta$ range. $Q_0$ shows a weak $\theta$ dependence, and has an average value of $1.8 \times 10^4$. $Q_0$ is mainly limited by the copper resistivity, and decreases as $\theta$ increases due to the slight cavity volume reduction. $C$ also shows a weak $\theta$ dependence, and has an average value of 0.62. The values of $C$ is calculated by a numerical integration of various factors according to Eq. \ref{eq:cavity form factor}. The result varies with the grid size slightly. The variation of $C$ due to different grid sizes is within 1\%, and is quoted as a systematic uncertainty in the analysis. The valuses of $f_{\rm c}$ and $C$ have a weak temperature dependence due to a slight contraction of the cavity volume at low temperature, while an increase of $Q_0$ by a factor of 3.5 at low temperature is expected from the copper resistivity reduction. The simulation suggests that the full $f_{\rm c}$ tuning range of 290 MHz is suitable for the axion search.

To apply a signal large enough for the cavity S-parameter measurements or the readout tests while not to introduce a significant cavity loss, the favored probe 1 coupling coefficient $\beta_1$ is around 0.01. To optimize the axion search, $\beta_2$ should be tuned around 2. Both probe coupling strengths $\kappa_i$ depend on their insertion depths to the cavity $d_i$. The relations are simulated. The comparisons of the simulation and the measurement results are discussed in Sec. \ref{subsec:cavity measurement}.

\subsubsection{Fabrication}

The cavity is made of the oxygen-free high conductivity copper from Hitachi Metals (part number: C1011), whose copper content exceeds 99.99\%. The material is first annealed under 500 \textdegree C for 15 minutes. The two halves of the split cavity are fabricated by the CNC machining. The inner corners are rounded with a radius of 3-mm in order to increase $Q_0$. The pieces undergo a surface treatment process, including polishing and chemical cleaning. Figure \ref{fig:cavity}(d) shows a photo of the split cavity. 

The off-axis tuning rod and its end connections to the teflon axles take a unibody design to increase $Q_0$. The copper material of the rod is not pre-annealed, because the mechanical strength of the pre-annealed copper is too weak for the rod with a large aspect ratio. Two teflon axles are installed to the rod ends. The top teflon axle is connected to the rotational motor through a copper rod.

A 50$\Omega$ semi-rigid coaxial cable (part number: JYEBAO .085CU-W-P-50) is used for the probe 2. The diameter of the inner conductor is 0.51 mm. The cable is stripped to let 6-mm long inner conductor uncovered. The cable is soldered to a SMA connector. The cable assembly is connected to the linear motor, and an adaptor is installed on the cavity top wall to guide the cable moving vertically. The linear motor can tune the probe insertion depth $d_2$, such that the coupling coefficient $\beta_2$ can be regulated during the experiment.

A standard SMA product (part number: JYEBAO SMA8640-0/7.55) with an inner conductor of 7.55 mm in length and 1.27 mm in diameter is used for the probe 1. A number of washers are used to set the probe insertion depth $d_1$, such that the coupling coefficient $\beta_1$ reaches the desired value. The size of the washer hole and the corresponding cavity hole is 2 mm in diameter to avoid extra modes induced by this additional space.

A RuO$_2$ thermometer is attached to the cavity directly to monitor the cavity temperature $T_{\rm c}$. Three copper bars of 283 mm in length support the cavity in the magnet bore center from the mixing flange, and provide ample thermal anchoring to cool the cavity temperature close to the DR base temperature $T_{\rm mx}$ = 27 mK. This mechanical installation of the cavity had been experimentally verified in the past to allow the cavity to reach the DR base temperature. However, in the CD102 run, $T_{\rm c}$ only reached 155 mK. The noise background from the cavity also suggested that $T_{\rm c}$ was at an elevated temperature. (See Fig. \ref{fig:IQtoPS}(d) for more details.) We believe that the cavity was not correctly isolated from the radiation shield.

The cavity and the components are kept in an acrylic vacuum chamber to prevent it from the surface oxidation when not being used.

\subsubsection{Motor}\label{subsec:Motor}

Two Attocube piezo-driven motors are utilized for cavity tuning: The rotational motor (ANR240) is to adjust the rod angle $\theta$ and hence the cavity frequency $f_{\rm c}$, and the linear motor (ANPz101eXT12) is to adjust the probe 2 insertion depth $d_2$ and hence the corresponding coupling quality factor $Q_2$. The rotational motor offers 360\textdegree~endless travel range, and connects directly to the frequency tuning rod. The linear motor offers 12 mm travel range, and connects directly to the readout probe 2 assembly with a weight load of 99 g. The motors are mounted on the mixing flange without any gear design.

The low-resistance copper dc wires in the DR connect the motors to the controller (ANC350). The controller drives the motors by a sawtooth waveform signal, and reads out the motor positions. The resistance values of the wires at room temperature for driving the rotational and linear motors are 6.7 $\Omega$ and 2.2 $\Omega$ (3 wire channels in parallel), respectively. These values approximately fulfill the suggested wire resistance of less than 5 $\Omega$, and the motors with their loads can move properly when the DR is at room temperature. The resistance values drop to 0.25 $\Omega$ and 0.083 $\Omega$, respectively, in the DR normal operation, and the motors were expected to work in this regard.

During the CD102 run, the rotational motor can function properly at the DR base temperature $T_{\rm mx}$ = 27 mK and $B_0$ = 8 T. However, the linear motor stopped moving during the cool down process as the system reached 70 K. We suspect that the connection bar had bent due to a thermal contraction. The bending caused higher friction between the probe 2 assembly and the adaptor, which stopped the linear motor from moving properly. We were therefore forced to use a fixed probe insertion depth $d_2$, which was preset at temperature higher than 70 K. However, it did not lead to a lower sensitivity in the CD102 run. Operational details that solved this problem is described in section Sec. \ref{subsec:Coupling study}. On the other hand, this issue may become critical for our future measurements and a more robust insertion mechanism to mitigate this problem is being studied.

The rotational motor moves via an open-loop operation in the axion search experiment. Each 60V sawtooth pulse drives a single rotational movement about 4 millidegrees, corresponding to a cavity frequency shift of 8 kHz. This shift value is smaller than the nominal frequency step $\Delta f_{\rm s}$ in the CD102 run (about 110 kHz). An approximately 14-pulse movement of the rotational motor is required for tuning a frequency step, causing an increase of the DR based temperature $T_{\rm mx}$ by few mK. The temperature rise forces us to wait for a few minutes after moving the rod for a frequency step. Meanwhile, we continuously monitor the motor positions. The two motor sensors consume less than 4 $\si{\micro \W}$ and does not raise the DR base temperature.

\subsubsection{Measurement}\label{subsec:cavity measurement}

\begin{figure*}[tb]
    \centering
    \includegraphics[width=0.7\textwidth]{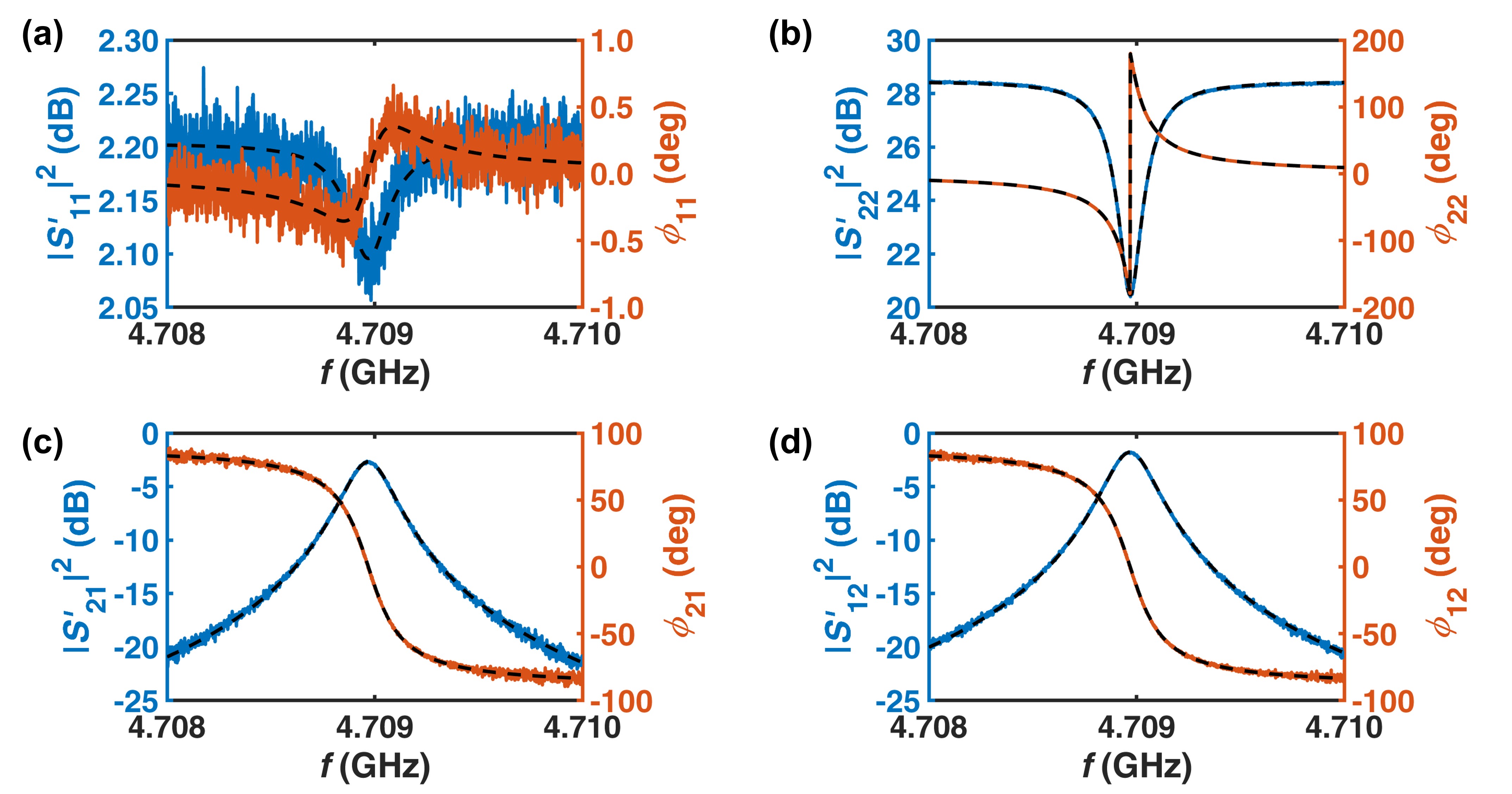}
    \caption{S-parameter measurements of cavity. (a) $S^{\prime}_{11}$, (b) $S^{\prime}_{22}$, (c) $S^{\prime}_{21}$, and (d) $S^{\prime}_{12}$. The blue and orange lines are the amplitude $|S|^2$ and the measured phase $\phi$, respectively. The dashed lines in (a) and (b) are the fitting curves. The fitting gives the cavity characteristics $f_{\rm c} = 4.70897\ {\rm GHz}$, $\kappa_0/2\pi = 70.63\ {\rm kHz}$, $\kappa_1/2\pi = 1.47\ {\rm kHz}$, and $\kappa_2/2\pi = 167.59\ {\rm kHz}$. The dashed lines in (c) and (d) are the derived curves from the cavity characteristics.}
    \label{fig:scattering matrix}
\end{figure*}

The elements of the scattering matrix of the two-port cavity near a resonance are
\begin{equation}
\label{eq:scattering matrix element}
\begin{aligned}
    S_{11} &= \frac{\left( \kappa_0-\kappa_1+\kappa_2 \right) + 2i\Delta}{\kappa + 2i\Delta}, \\
    S_{22} &= \frac{\left( \kappa_0+\kappa_1-\kappa_2 \right) + 2i\Delta}{\kappa + 2i\Delta}, \\
    S_{21} &= S_{12} =\frac{2\sqrt{\kappa_1\kappa_2}}{\kappa + 2i\Delta},
\end{aligned}
\end{equation}
where the detuning $\Delta = 2\pi(f-f_{\rm c})$. Note that in a practical S-parameter measurement, the baseline is affected by the attenuation of the MW input line and the gain of the output amplification line. Therefore, the measured value, $S^{\prime}_{mn}$, and the true value, $S_{mn}$, in a small frequency scan range are different by a frequency-independent factor due to the overall gain of the measurement wiring. Without the knowledge of the attenuation/gain of the input/output lines, fitting $S^{\prime}_{11}$ and $S^{\prime}_{22}$ together to Eq. \ref{eq:scattering matrix element} is enough to precisely determine all the cavity characteristics $f_{\rm c}$, $Q_0$, $Q_1$, and $Q_2$. Figure \ref{fig:scattering matrix} shows an example. The S-parameter measurements are taken as the cavity is installed in the DR with the DR MW wiring; see Sec. \ref{subsec:Microwave layout} and Fig. \ref{fig:MWwiring} for more details about the experimental setup for the S-parameter measurements performed in the DR. The dashed lines in Fig. \ref{fig:scattering matrix}(a)(b) show the fitting results of $S^{\prime}_{11}$ and $S^{\prime}_{22}$, and give $f_{\rm c}$ = 4.70897 GHz, $Q_0 = 2\pi f_{\rm c}/\kappa_0 = 6.67 \times 10^4$, $Q_1 = 3.20 \times 10^6$, and $Q_2 = 2.81 \times 10^4$. With the obtained cavity characteristics, $S_{21}$ and $S_{12}$ are calculated, and displayed as the dashed lines in Fig. \ref{fig:scattering matrix}(c)(d) with the overall factors from the gains of the measurement wiring, respectively. The consistency of the measured $S^{\prime}_{21}$, $S^{\prime}_{12}$ and calculated $S_{21}$ and $S_{12}$ indicates the quality of the comprehension of the cavity characteristics. Note that fitting to solely the $S^{\prime}_{22}$ measurement can determine the cavity characteristics $f_{\rm c}$, $Q_2$, and $Q_{01} = 2\pi f_{\rm c}/(\kappa_0 + \kappa_1) \approx Q_0$, and therefore $\eta$. 

\begin{figure}[b]
    \centering
    \includegraphics[width=0.48\textwidth]{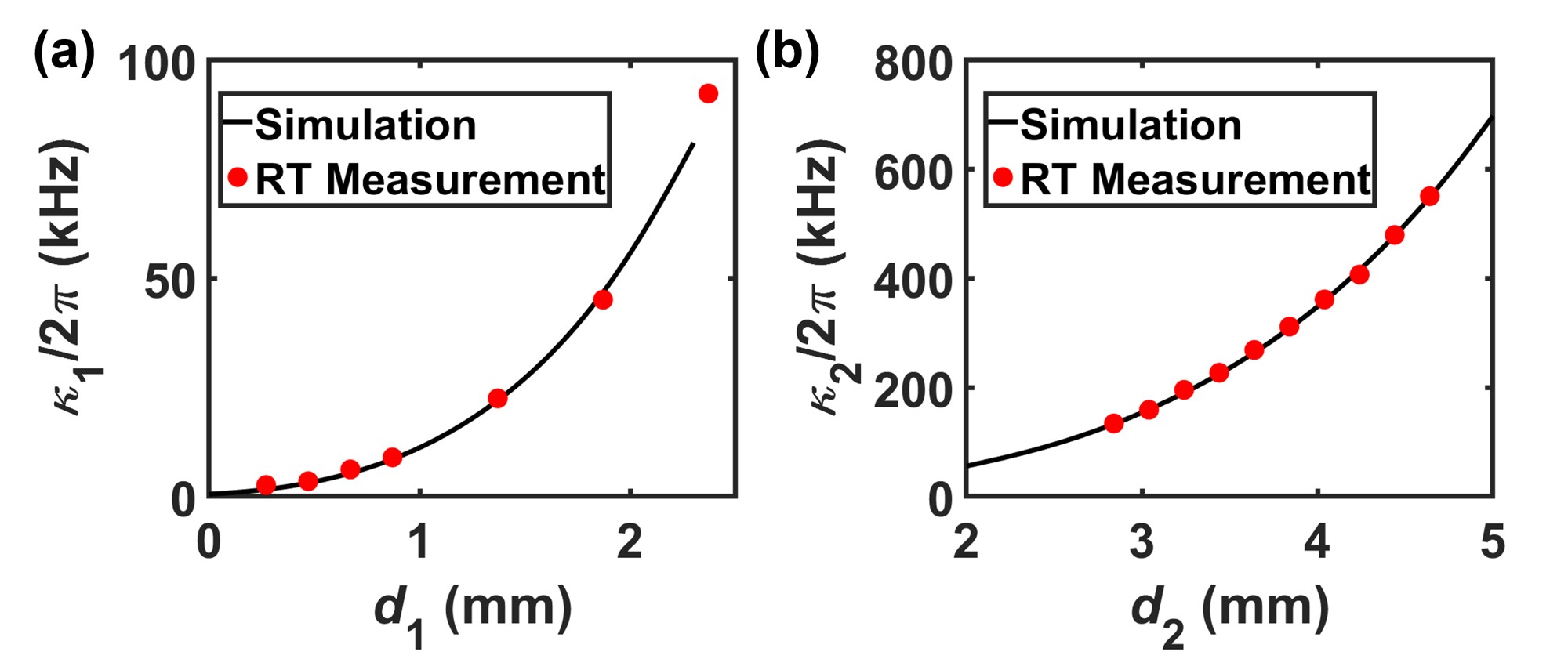}
    \caption{Cavity probe coupling study of (a) probe 1 and (b) probe 2 at room temperature. Both simulation (black line) and experiment (red symbol) results are presented. The study is at $\theta$ = 85\textdegree. Spline fitting is used to display the simulation result. For better comparison, the experimentally measured $d_1$ and $d_2$ are shifted horizontally with respect to the simulated results by -0.13 and 0.64 mm, respectively.}
    \label{fig:probe test}
\end{figure}

Via the S-parameter measurements, an experimental study of the coupling strength $\kappa_{1,2}$ vs. the insertion depth $d_{1,2}$ for both probes at room temperature is performed. Figure \ref{fig:probe test} shows the experimental results, accompanied with the HFSS simulation results for comparison. The study is at $\theta$ = 85\textdegree. The $\kappa_1$ study focuses on $\beta_1$ around 0.01, and the $\kappa_2$ study focuses on $\beta_2$ in the range of $1-3$. (The simulation and the experiment $\kappa_0/2\pi$ are 250 kHz and 264 kHz, respectively, to determine $\beta$.) The experiment and simulation results agree well with each other for both probes. (For better comparison, the experimentally measured $d_1$ and $d_2$ are shifted horizontally with respect to the simulated results by -0.13 and 0.64 mm, respectively. The shifts are consistent with the mechanical measurement uncertainty of the absolute insertion depths, $\approx$ 0.5 mm.) The values of $\kappa_{1,2}$, and therefore $\beta_{1,2}$, increase with $d_{1,2}$. The agreements illustrate that we have good controls of the effect of varying the insertion depths to establish the corresponding $\beta$ values. Note that the relations of $\kappa_{1,2}$ vs. $d_{1,2}$ are rather temperature independent. Based on the coupling study results and an expected $Q_0$ increase from the copper resistivity reduction at low temperature, $d_1$ is set at 0.4 mm at room temperature to roughly achieve $\beta_1$ = 0.01 in the cryogenic condition. $\beta_1 \approx 0.02$ is experimentally realized at the DR base temperature. The probe 2 is installed with the linear motor such that the motor movement can assist $\beta_2$ to cover the range of $1-3$ comfortably in the cryogenic condition.

\begin{figure*}[tb]
    \centering
    \includegraphics[width=0.7\textwidth]{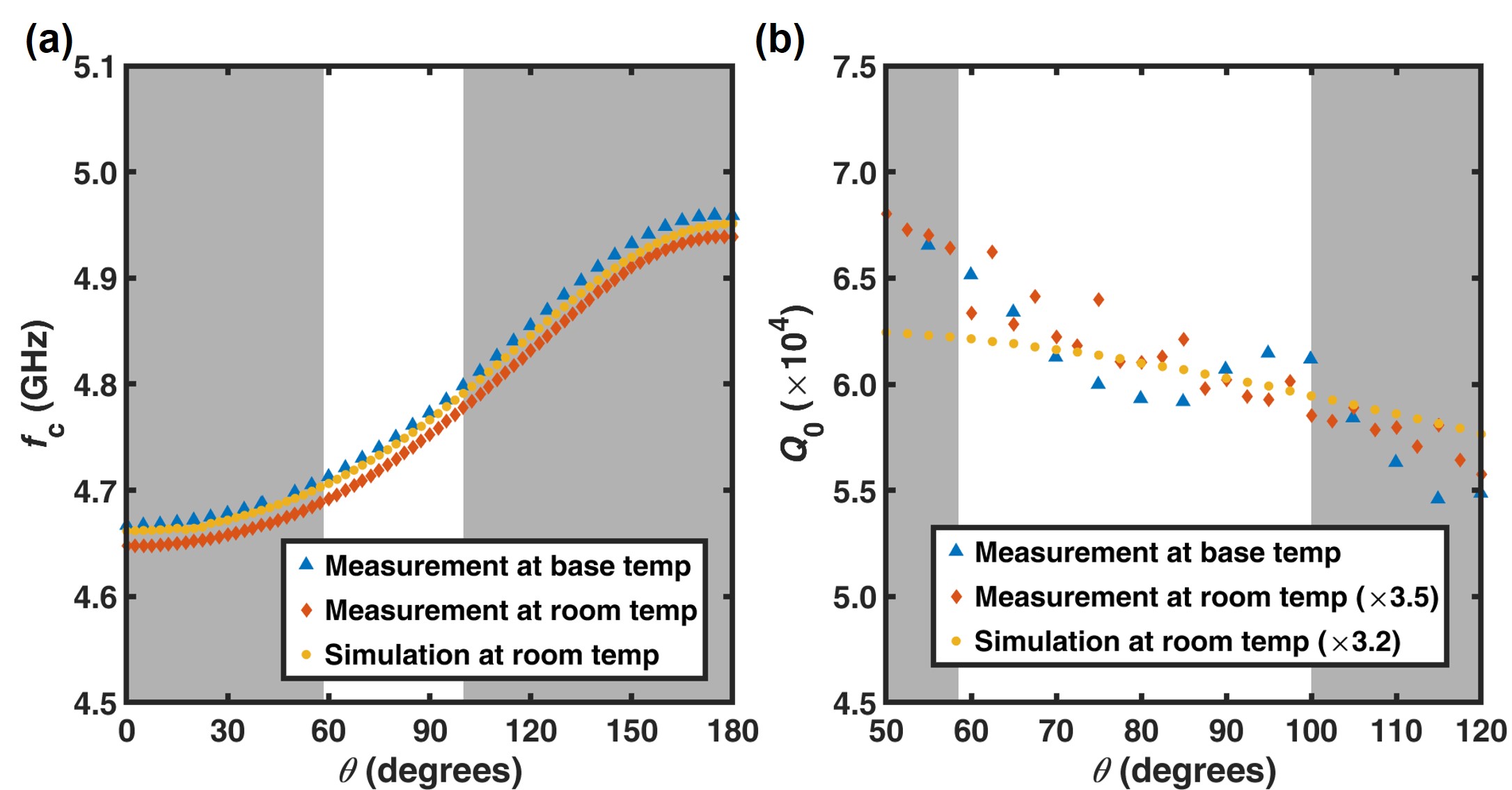}
    \caption{Comparisons of TM$_{010}$ mode (a) resonant frequency $f_{\rm c}$ and (b) intrinsic quality factor $Q_0$. The yellow dots are the room temperature simulation results. The orange diamonds and blue triangles are derived from fitting the experimental data taken at room temperature and at the DR base temperature, respectively. Note that in (b) the $Q_0$ of the room temperature simulation and experimental results are multiplied by a factor of 3.2 and 3.5, respectively, to have the comparison more straightforward. The axion search range in the CD102 run is indicated by the white ground.}
    \label{fig:Resonant Frequency Comparison}
\end{figure*}

The cavity characteristics vs. the tuning bar position $\theta$ are determined through the $S^{\prime}_{22}$ measurement. Figure \ref{fig:Resonant Frequency Comparison} shows the results of the TM$_{010}$ mode frequency $f_{\rm c}$ in (a), and the intrinsic quality factor $Q_0$ in (b). The orange diamonds represent the results at room temperature. The blue triangles represent the results at $T_{\rm c}$ = 155 mK, which appears at the DR operational condition of $T_{\rm mx}$ = 27 mK. The room temperature simulation (yellow dots) is included for comparison. The three data sets of $f_{\rm c}$ in (a) have nearly identical trends. Most importantly, the data show no mode crossing. The room temperature data depict resonant frequencies $\approx 14$ MHz lower than those predicted by the simulation, indicating that the cavity inner diameter $D$ is 0.29\% larger than the designed value. The DR base temperature data depict overall 20 MHz higher resonant frequencies than the corresponding room temperature ones, indicating a 0.42\% thermal contraction of $D$, consistent with the expectation from the copper thermal expansion coefficient. 

The three data sets of $Q_0$ in (b) also have similar trends in the 50\textdegree~$-$ 120\textdegree~$\theta$ range. Note that $Q_0$ of the room temperature simulation and experimental results are multiplied by a factor of 3.2 and 3.5, respectively, to have the comparison more straightforward. The effective surface-to-volume ratio increases as $\theta$ increases, causing $Q_0$ decreases accordingly. The average value of $Q_0 = 1.8 \times 10^4$ extracted from the room temperature $S^{\prime}_{22}$ data is approximately 9\% lower than the simulation one using the bulk copper resistivity, likely due to the imperfect surface condition of the cavity. $Q_0$ reaches an average value of $6.1 \times 10^4$ at the DR base temperature, compatible with the temperature dependence of the copper resistivity. $Q_0$ at the DR base temperature has an oscillation structure with a relative amplitude of 5\%. The cause of the phenomenon is unknown to us, but will solely affect the search limit by about 1\% if the uncertainty of the derived $Q_0$ is on the same order.

The axion search range in the CD102 run is indicated by the white ground in Fig. \ref{fig:Resonant Frequency Comparison}. The comparisons of the measurement and simulation studies conclude that the simulations of the TM$_{010}$ mode properties are reliable. This conclusion is particularly important to establish the confidence of the simulated form factor $C$ used in the data analysis.

In addition to the expected increase due to the lower copper resistivity at the DR base temperature, $Q_0$ increases slightly during the magnetic field ramp-up. The effect saturates to a 5.2\% increase at around $B$ = 3 T. The increase favors our axion search even though the physics of the phenomenon is unknown to us and under investigation.

\subsection{Microwave system}\label{subsec:Microwave system}

A MW system is used to examine the cavity characteristics, and to transmit and to record the signal to and from the cavity. Figure \ref{fig:MWwiring} shows the schematic of the MW design. The TASEH MW detection system is prepared to perform the S-parameter measurements, the amplification chain calibration, and the axion detection experiment.

\subsubsection{Microwave layout}\label{subsec:Microwave layout}

\begin{figure}[tb]
    \centering
    \includegraphics[width=0.48\textwidth]{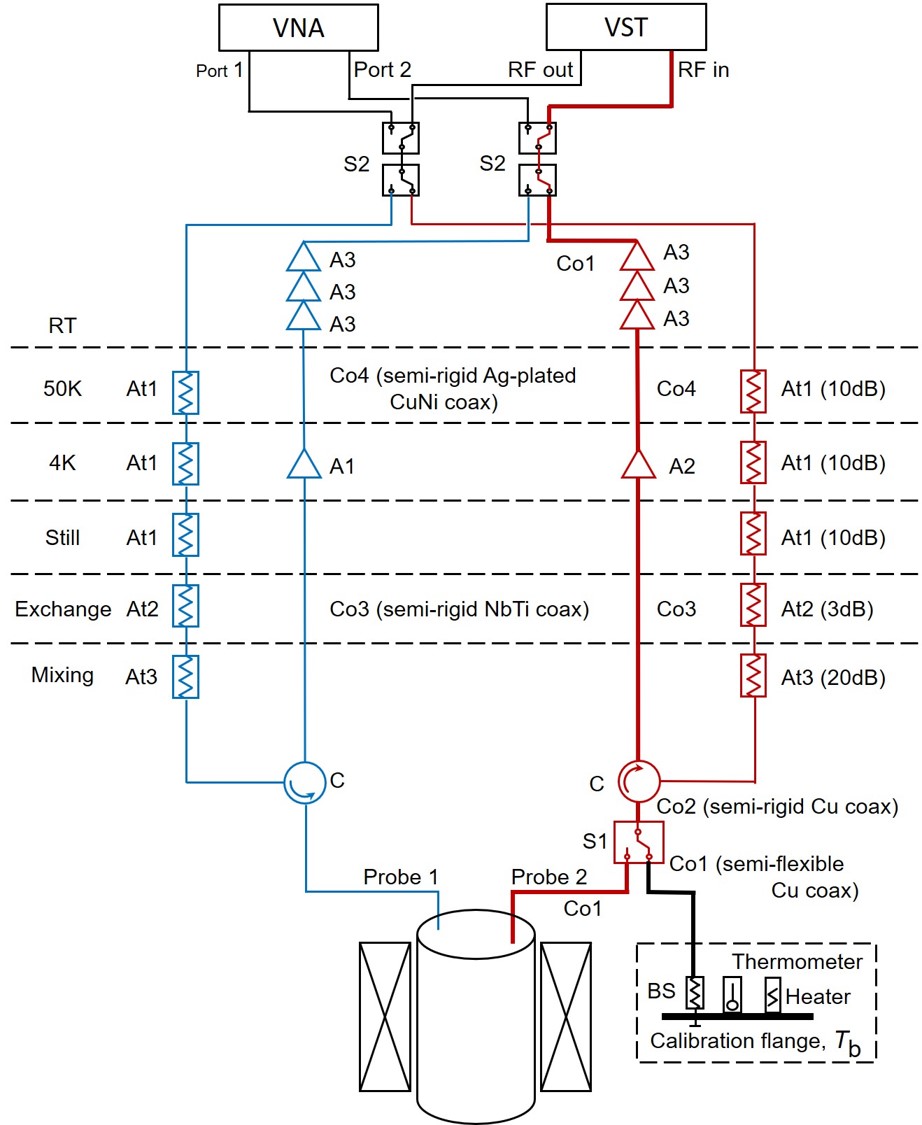}
    \caption{Schematic diagram of TASEH MW design. The dashed horizontal lines separate the DR flanges. The MW system mainly consists of two sets of input/output lines. Each set connects to one cavity probe through a circulator to allow the full S-parameter measurements of the two-port cavity. The lines associated to the probes 1 and 2 are marked by the blue and red colors, respectively. The cavity readout probe 2 and the thick output line are responsible for the axion data taking. Via a cryogenic MW switch, this output line can also connect to the blackbody radiation source (BS), made of a $50\Omega$ terminator, for calibration. The VNA performs the S-parameter measurements of the two-port cavity. The VST carries out the signal acquisition and/or generates the test signals to the system. The descriptions, such as coax types, and the part numbers of the labeled items are listed in Tab. \ref{tab:component} in Appendix \ref{sec:Table}.}
    \label{fig:MWwiring}
\end{figure}

A high-gain, low-noise amplification chain, named amplification chain 2 and marked by the thick lines in Fig. \ref{fig:MWwiring}, is used for the axion detection. The first stage amplifier for this chain is a LNF-LNC4\_8C low-noise HEMT amplifier (A2) anchored at the 4K flange. This amplifier offers approximately 40 dB gain and 1.5 K noise temperature in the $4-8$ GHz frequency span. The following three amplifiers at room temperature provide 20 dB gain each to boost the signal strength for follow-up processing. A LNF-CIISISC4\_8A three-stage circulator (C), providing approximately 60 dB isolation but also leading to 0.4 dB insertion loss, is anchored at the mixing flange to prevent thermal radiation from the HEMT amplifier from back streaming to the cold cavity and then reflected by the cavity. Several low-loss coaxial cables guide the signal through the components, including a superconducting cable between the circulator and the HEMT amplifier. The signal from the cavity probe 2 is routed to this amplification chain. Knowing the overall gain $G_2$ and the added noise $T_{\rm a2}$ of the amplification chain 2 is essential to derive the output power at the cavity probe 2 from $P_2$, the readout power in the signal receiver, and to obtain the SNR of the axion search. $G_2 \approx$ 100 dB and $T_{\rm a2} \approx$ 2 K are obtained from the calibration results discussed in Sec. \ref {subsec:Amplification calibration}.

One more supplementary output amplification chain 1 and two heavily attenuated input lines 1 and 2, connected to the probes 1 and 2, respectively, are to support the cavity S-parameter measurements. The lines associated to the probes 1 and 2 are marked by the blue and red colors, respectively. The amplification chain 1 has similar setting and characteristics as the data taking amplification chain 2. The two attenuated lines, used to input MW signals, have one attenuator thermally anchored at each cold flange to reduce the broadband radiation from the higher-temperature environment or flanges. Each circulator connects one cavity probe to one input line and one output amplification chain to permit reflection-type measurements. Although those chains are designed for cryogenic measurements, they do allow the S-parameter measurements to examine the cavity characteristics at room temperature. The input line 1 also delivers the synthetic axion signal.

Two major MW instruments are used in the TASEH experiments. A vector network analyzer (VNA) performs the S-parameter measurements of the two-port cavity. A vector signal transceiver (VST), comprising a vector signal analyzer and a vector signal generator, conducts the signal acquisition and is also responsible for generating the synthetic axion signals to the system. A $50\Omega$ terminator severs as a broadband blackbody radiation source (BS) to calibrate the data taking amplification chain. A copper plate, called calibration flange and mounted on the mixing flange via two stainless steel pillars as weak thermal links for cooling, provides the thermal bath for the radiation source. The radiation source temperature $T_{\rm b}$ can be controlled and monitored through a 220$\Omega$ resistor heater and a cernox thermometer on the plate. A cryogenic MW switch (S1) directs the signal from either the cavity or the BS to the data taking amplification chain 2. Four room-temperature MW switches (S2) manage the signal flows for various types of measurements. 

\subsubsection{Signal acquisition and generation}\label{SigAcq_Gen}

The MW signal acquisition is relied on the vector signal analyzer part of the National Instruments PXIe-5644R VST. The signal from the cavity is amplified and directed to the VST input for the data acquisition. The VST performs IQ demodulation and sampling. The sampled IQ quadrature time trace data are sent to a network attached storage (NAS). The fast Fourier transform (FFT) algorithm is performed offline in real time to retrieve the frequency-domain power spectrum.

The VST downconverts the incoming MW signal with respect to the local oscillator (LO) frequency $f_{\rm LO}$ to the baseband IQ quadrature signals. The frequency accuracy is $\pm 2.2 \times 10^{-6}$~\cite{NIPXIe5644r2017manual}. It has 16-bit analog-to-digital converters (ADCs) to digitize the IQ signals. The IQ sampling rate $F_{\rm s}$ = 2 MHz is set. IQ data of 1 second are saved as a single file, which takes 32 MB storage space in the NAS. The overall data acquisition time efficiency is 98.4\%.

An acquisition time $t_{\rm a}$ = 1 ms is chosen to give IQ time traces of 2000 points. The evaluated FFT signal power spectrum has a 2-MHz frequency bandwidth centered at $f_{\rm LO}$ and has the corresponding spectral resolution bandwidth $\Delta f = 1/t_{\rm a}$ = 1 kHz, small enough to resolve an expected axion signal of 5 kHz width. (In this paper most of the frequencies in unit of GHz quote 6 decimal places as the resolution bandwidth is 1 kHz. It should be noted that the absolute accuracy of the frequency is around 10 kHz.) To avoid the aliasing effect, the VST rolls off the IF signal by a band-pass filter, imposing the useful frequency span of $0.8F_s$ = 1.6 MHz. The span is large enough to cover the cavity bandwidth $\Delta f_{\rm c}$ around $f_{\rm c}$ in the CD102 run (about 240 kHz), the sensitive frequency range to the axion converted signal.

\begin{figure*}[tb]
    \centering
    \includegraphics[width=0.7\textwidth]{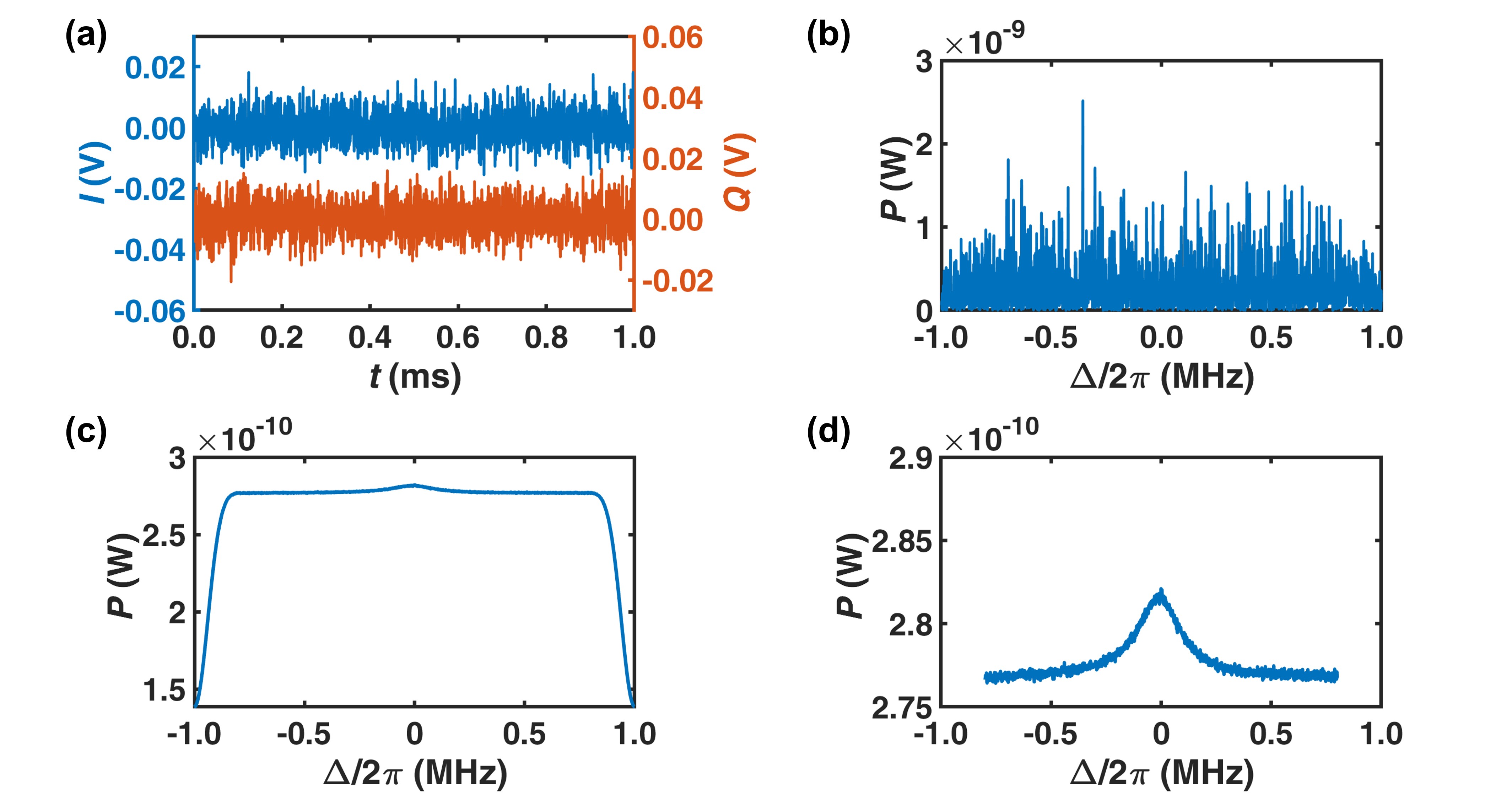}
    \caption{Demonstration of the signal acquisition. (a) 1-ms IQ time trace data with $f_{\rm LO} = f_{\rm c}$ = 4.708970 GHz. (b) Power spectrum of the 1-ms IQ data in (a). (c) Average power spectrum of $t$ = 60 minute. (d) Zoom-in of the useful span in (c).}
    \label{fig:IQtoPS}
\end{figure*}

Figure \ref{fig:IQtoPS} demonstrates a signal acquisition from the cavity in the DR with $f_{\rm c}$ = 4.708970 GHz. $f_{\rm LO} = f_{\rm c}$ is set for the acquisition. The spectrum is dominated by the background noise of the cavity plus the added noise of the amplification chain. Figure \ref{fig:IQtoPS}(a) shows 1-ms IQ time trace data, and (b) is the power spectrum of the IQ data in (a), evaluated via FFT. (c) is the power spectrum averaged over an integration time $t$ = 60 minute, corresponding to the number of spectra $N=3.6 \times 10^6$ for taking the average. The fluctuation of the average noise power is reduced by a factor of $\sqrt{N}$. The roll-offs at the edges due to the IF filtering are clearly seen. (d) is the zoom-in of the useful span in (c). The spectrum baseline is from the amplification chain added noise plus the vacuum noise, which is expected to be white. The Lorentzian-like excess noise in the middle originates from the cavity thermal noise. The behavior is consistent with that the cavity is at an elevated temperature $T_{\rm c}$ = 155 mK measured by the attached RuO$_2$ thermometer and the cavity bandwidth $\Delta f_{\rm c}$ = 196 kHz~\cite{khatiwada2021axion}.

The vector signal generator part of the NI PXIe-5644R VST can generate a synthetic MW signal up to 6 GHz with an arbitrary lineshape via IQ modulation. The VST takes 16-bit designed IQ time sequences and has the maximal updating rate of 120 MHz. The modulation bandwidth is 80 MHz, much larger than the target test signal width of 5 kHz. A test experiment with a synthetic signal to verify the data acquisition is discussed in Sec. \ref{subsec:Synthetic axion experiment}. 

\subsubsection{Amplification calibration and environment monitoring} \label{subsec:Amplification calibration}

\begin{figure*}[tb]
    \centering
    \includegraphics[width=0.7\textwidth]{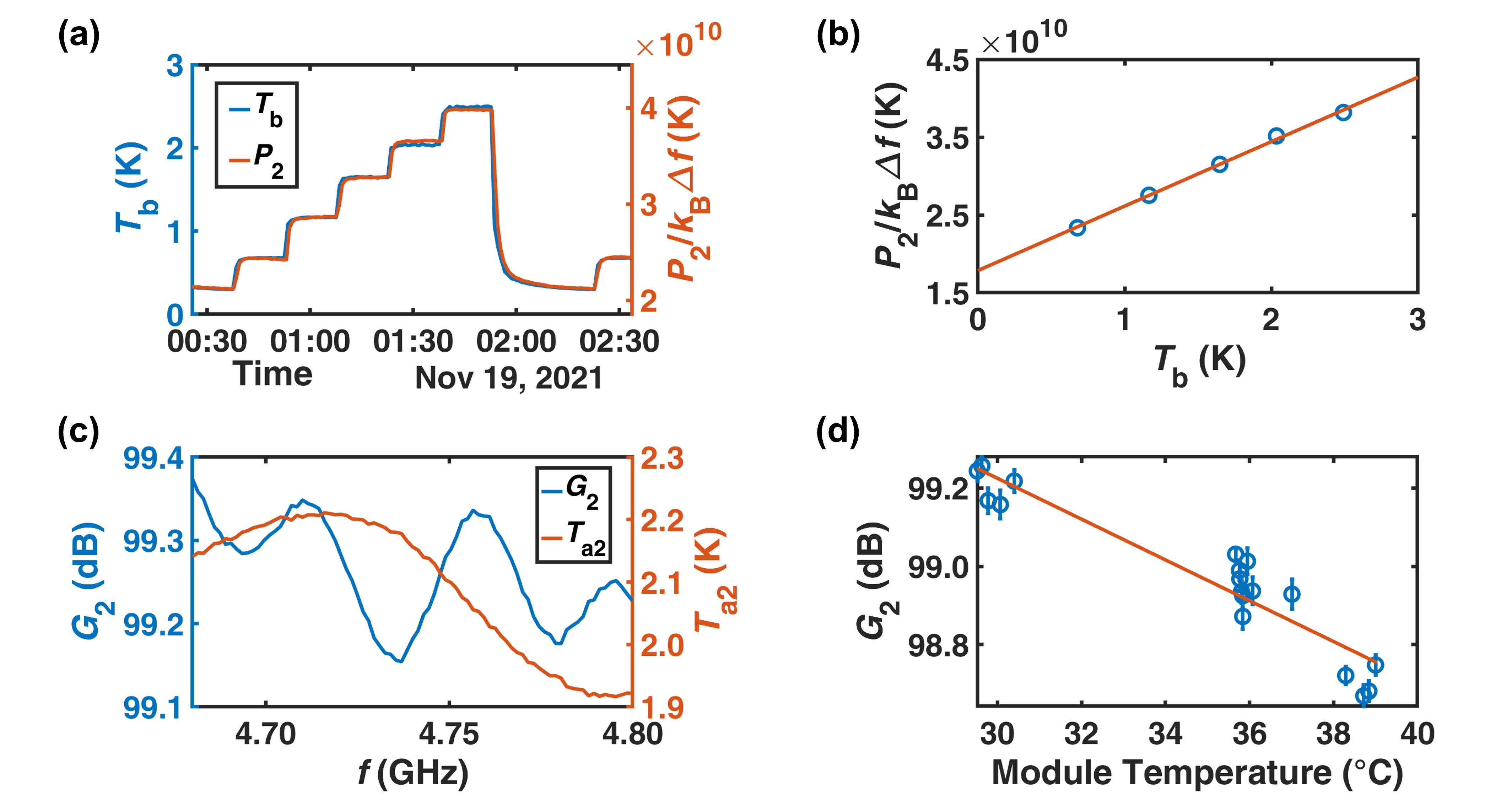}
    \caption{Calibration of the amplification chain. (a) Time trace records of $P_2$ and $T_{\rm b}$ at $f$ = 4.68 GHz. (b) $P_2$ vs. steady-state $T_{\rm b}$. The blue open circles represent the data from (a), and the red line is the linear fitting. (c) $G_2$ and $T_{\rm a2}$ vs. $f$. (d) $G_2$ at $f = 4.68\ {\rm GHz}$ as a function of the VST module temperature. The red line is a linear fitting.}
    \label{fig:HEMTcal}
\end{figure*}

As described in Sec. \ref{subsec:Microwave layout}, the BS with the controlled temperature $T_{\rm b}$ is used to confirm the gain $G_2$ and the added noise $T_{\rm a2}$ of the amplification chain 2 for axion data taking via the Y-factor method. As the cryogenic switch directs the power of the BS via the amplification chain 2 to the VST, in this case the readout power $P_2$ in a frequency bin at the frequency $f$ is
\begin{equation}
\label{eq:BSpower}
    P_2 = G_2 \left[ hf \left( \frac{1}{e^{hf/k_{\rm B}T_{\rm b}} - 1} +\frac{1}{2} \right) + k_{\rm B}T_{\rm a2} \right]\Delta f,
\end{equation}
where the first and the second terms are the noises from the BS, and the third term is from the amplification chain. As $T_{\rm b} \gg hf/k_{\rm B}$ (about 0.24 K for $f \approx$ 5 GHz), Eq. \ref{eq:BSpower} reduces to 
\begin{equation}
\label{eq:BSpowerlinear}
    P_{2} \approx G_2 k_{\rm B} \left[ T_{\rm b} + T_{\rm a2} \right]\Delta f.
\end{equation}
By varying $T_{\rm b}$ above $hf/k_{\rm B}$ through the 220$\Omega$ heater and fitting the linear dependence of $P_2$ vs. $T_{\rm b}$, $G_2$ and $T_{\rm a2}$ can be calibrated. Figure \ref{fig:HEMTcal} shows the calibration data and results. As a current flows through the heater in steps of 0.3 mA up to 1.5 mA, 15 minutes for each step, $T_{\rm b}$ raises to 5 steady temperatures accordingly. Figure \ref{fig:HEMTcal}(a) shows both $P_2$ at $f$ = 4.68 GHz and $T_{\rm b}$ as functions of time. Indeed $P_2$ follows $T_{\rm b}$ well as $T_{\rm b}$ varies. Figure \ref{fig:HEMTcal}(b) depicts the linear relation between $P_2$ and the average of the steady-state $T_{\rm b}$. By means of linear fitting $G_2$ = 99.37 dB and $T_{\rm a2}$ = 2.14 K are confirmed. Figure \ref{fig:HEMTcal}(c) shows the frequency dependence of $G_2$ and $T_{\rm a2}$ in the $4.68-4.80$ GHz range. $T_{\rm a2}$ is about $1.9 - 2.2$ K in this frequency range; the value is reasonable when taking into account the nominal 1.5 K noise temperature of the HEMT amplifier plus approximately overall 1.0-1.7 dB attenuation (from the cables, switch, and circulator) before the HEMT. Together with the cavity at the elevated temperature $T_{\rm c}$ = 155 mK, the value of the system noise of the detection $T_{\rm sys} = \Tilde{T}_{\rm c} + T_{\rm a2}$ is about $2.1 - 2.4$ K, roughly a factor of $9 - 10$ above the quantum-limited performance ($T_{\rm sys} = hf/k_{\rm B}$ = 0.24 K). This frequency-dependent $T_{\rm sys}$ is used for axion search data analysis~\cite{TASEH2022analysis}. Besides, the weak oscillations of $G_2$ vs. $f$ with an amplitude on the order of 0.1 dB could be owing to the impedance mismatch of a particular cable in this amplification chain. 

We notice that $P_2$ drifts over time. A monitoring system is built to discover how different environment parameters affect $G_2$ and/or $T_{\rm a2}$. The system tracks the DR flange temperatures, DR circulation and vacuum space pressures, cooling water temperature, lab temperature and humidity, VST module temperature, HEMT bias voltages, and 110V power line voltage. The correlation analysis unfolds that solely the VST module temperature influences $G_2$, as shown in Fig. \ref{fig:HEMTcal}(d). $T_{\rm a2}$ is fairly insensitive to the environment parameters. The frequency-dependent system noise temperature $T_{\rm sys}$ obtained from the calibrated added noise $T_{\rm a2}$ in Fig. \ref{fig:HEMTcal}(c) plus the $\Tilde{T}_{\rm c}$ of $T_{\rm c}$ = 155 mK is used for data analysis~\cite{TASEH2022analysis}. Meanwhile, the $G_2$ calibrated by the time-dependent VST module temperature was used to estimate the system noise directly from the physical data during the CD102 run via $T_{\rm sys} = P_2/G_2 k_{\rm B} \Delta f$. The estimated and the calibrated $T_{\rm sys}$ agreed very well~\cite{TASEH2022analysis}.
\section{Experimental procedure}

From Eq. \ref{eq:axion_signal_power} and the optimal experimental parameters of the current TASEH setup, the expected signal power is $P_{\rm a} \approx 1.4 \times 10^{-24}$ W for a benchmark KSVZ axion with a mass of $19.6\ \si{\micro \eV}$. The system noise power in a 5-kHz expected axion signal width is $P_{\rm n} = k_{\rm B} T_{\rm sys} \Delta f \approx 1.6 \times 10^{-19}$ W. The search goal of the CD102 run is to detect an axion converted signal with $g_{\rm a\gamma\gamma}$ a factor 10 of the KSVZ model. This section discusses the choices of the operation parameters and the experiment procedures. A synthetic axion experiment to verify the procedures of data acquisition and physics analysis is also described. Some unexpected experiment issues are discussed at the end.

\subsection{Coupling dependence of detection limit}\label{subsec:Coupling study}

Due to the issue of the linear motor as described in Sec. \ref{subsec:Motor}, the coupling coefficient of the probe 2, $\beta_2$, cannot be adjusted during the data taking at the DR base temperature. The coupling dependence of the detection limit is studied to understand the damage of the issue.

For an axion haloscope with certain system performances, the operation settings, including the cavity frequency shift step $\Delta f_{\rm s}$, detection probe coupling coefficient $\beta_2$, and integration time $t$, determine the search limit and range~\cite{al2017design}. From Eqs. \ref{eq:axion_signal_power} and \ref{eq:Dicke_eq}, one can see SNR $\propto g_{\rm a\gamma\gamma}^2 \beta_2/(1+\beta_2)^2 t^{1/2}$. For a fixed $g_{\rm a\gamma\gamma}$ search limit and SNR, $t \propto (1+\beta_2)^4/\beta_2^2$. In addition, only within the cavity bandwidth $\Delta f_{\rm c} = \kappa/2\pi \propto (1+\beta_2)$ around $f_{\rm c}$ is sensitive to the axion converted signal. To have proper overlaps of the sensitive regions such that the frequency scan is continuous, the frequency shift step $\Delta f_{\rm s} \lesssim \Delta f_{\rm c}/2$ is adopted. Therefore, for a fixed $g_{\rm a\gamma\gamma}$ search limit, the scan rate $\Delta f_{\rm s}/t$ is proportional to $\beta_2^2/(1+\beta_2)^3$. Consequently, $\beta_2 = 2$ optimizes the scan. 

\begin{figure*}[tb]
    \centering
    \includegraphics[width=0.86\textwidth]{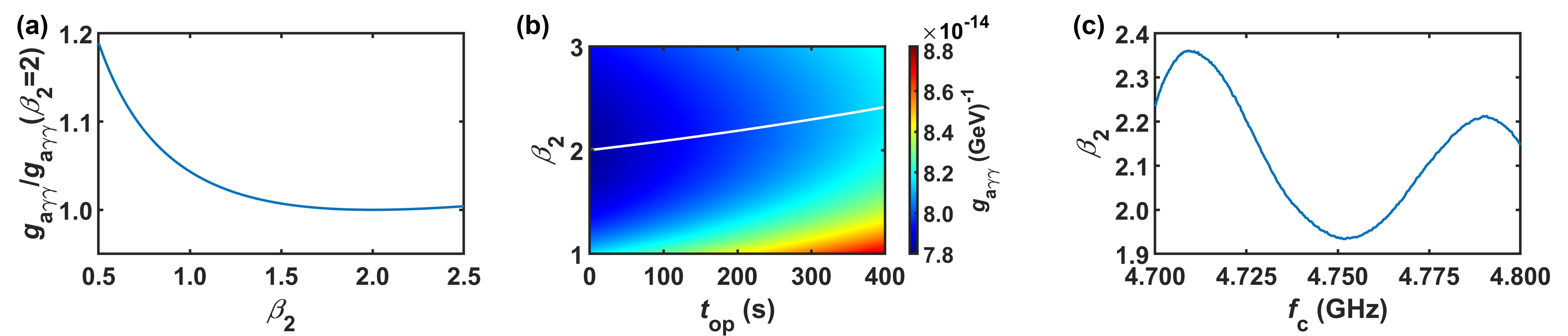}
    \caption{Coupling dependence of detection limit. (a) $g_{\rm a\gamma\gamma}/g_{\rm a\gamma\gamma}(\beta_2=2)$ vs. $\beta_2$ for a fixed scan rate. (b) $g_{\rm a\gamma\gamma}$ limit vs. $\beta_2$ and $t_{\rm op}$. The white line indicates the optimal $\beta_2$ values at various $t_{\rm op}$. (c) Experimental $\beta_2$ values vs. $f_{\rm c}$ at the DR base temperature with $d_2$ = 2.3 mm.}
    \label{fig:beta study}
\end{figure*}

On the other hand, for a fixed scan rate, $g_{\rm a\gamma\gamma}$ limit is proportional to $((1+\beta_2)^3/\beta_2^2)^{1/4}$. Figure \ref{fig:beta study}(a) shows $g_{\rm a\gamma\gamma}$ limit vs. $\beta_2$. The best limit again occurs at $\beta_2 = 2$. Yet in the $1.6 < \beta_2 < 2.5$ range, the limit gets worse only by less than 1\% compared with the optimal $\beta_2$ case. 

Furthermore, an operation time $t_{\rm op}$ for the actions during each cavity frequency shift is required. In our current setup $t_{\rm op} \approx$ 5 minute (see Sec. \ref{subsec:axionExp and rescan} for more details). The effect of $t_{\rm op}$ in determining the scan rate is studied. Instead of considering the integration time $t$ alone, the overall time $t + t_{\rm op}$ is used to determine the scan rate. Figure \ref{fig:beta study}(b) shows $g_{\rm a\gamma\gamma}$ limit vs. $\beta_2$ and $t_{\rm op}$. The study uses a set of parameters ($B_0$ = 7.8 T, $V$ = 0.234 liter, $C$ = 0.67, $Q_0$ = 60000, $f$ = 4.7 GHz) similar to our haloscope system performances. The optimal $\beta_2$ increases from 2 gradually as $t_{\rm op}$ increases. When $t_{\rm op} \approx$ 5 minute, the optimal $\beta_2$ is about 2.3, and the $1.9 < \beta_2 < 2.8$ range gives the limit within a variation around 1\%. The insensitive coupling dependence of the detection limit near the optimal $\beta_2$ encouraged us to ignore the linear-motor issue at the DR based temperature in the CD102 run, and preset $d_2$ when the linear motor was movable at higher temperature to obtain $1.9 < \beta_2 < 2.8$ for the axion search. Figure \ref{fig:beta study}(c) shows the experimental $\beta_2$ values at the DR base temperature with the preset $d_2$ = 2.3 mm. In the search frequency range of $4.7 - 4.8$ GHz, the variation of $\beta_2$ is within $1.9 - 2.4$, right inside the target range. Although fixed $d_2$ worked successfully for the CD102 data taking, we still need a linear motor to control the $d_2$ for future experiment operations, for rescan is most efficient at $\beta_2$ = 1, and $\beta_2$ with fixed $d_2$ may go out of range when the cavity frequency tuning range is very wide.

Note that as the required integration time $t$ gets short in the future upgraded experiments, $t_{\rm op}$ could become significant. Consequently, the optimal $\beta_2$ deviates from 2 notably, and the effect should be taken into account when designing efficient data taking strategies or procedures.  

\subsection{Axion detection and rescan procedure} \label{subsec:axionExp and rescan}

The goal of our current search is to detect an axion converted signal with $g_{\rm a\gamma\gamma} = 7.4 \times 10^{-14}\ {\rm GeV}^{-1}$, 10 times above the KSVZ model value. To fulfill $\Delta f_{\rm s} \lesssim \Delta f_{\rm c}/2 \approx 120\ {\rm kHz}$, a nominal value of $\Delta f_{\rm s}$ = 105 kHz is chosen, and a 10\% tolerance of $\Delta f_{\rm s}$ is allowed. According to our current haloscope performances, and taking frequency dependent $T_{\rm sys}$, $Q_0$, and $\beta_2$ into account, the required $t = 32 - 42$ minute is implemented in each cavity frequency tuning step. Such integration time $t$ corresponds to the number of spectra for taking the average $N = 1.92 \times 10^6 - 2.52 \times 10^6$. 

Each cavity frequency tuning step takes three moves for the axion detection:
\begin{enumerate}
\item Drive the rotational motor with a signal of 14 sawtooth pulses for a target cavity-tuning-rod movement of 56 millidegrees.
\item Perform a $S^{\prime}_{22}$ measurement to confirm $\Delta f_{\rm s}$ within the $95 - 115$ kHz range. If not, tune the rod angle again to achieve the required $\Delta f_{\rm s}$. 
\item Set VST $f_{\rm LO}$ to the updated $f_{\rm c}$, and begin a data acquisition for a period of $t$.
\end{enumerate}
The procedure repeats to scan $f_{\rm c}$ over the intended frequency range. The operation time of those procedures takes about 5 minutes. The DR temperature typically rises by a few mK due to the rotational motor movement. By the time the data acquisition starts, the DR temperature drops back to the base temperature $T_{\rm mx}$ = 27 mK. Therefore the temperature rise does not affect the data taking.

The frequency bins containing the axion converted signal, or other interference signals, will possess higher power. However, due to the system noise from the cavity and the signal receiver, some random bins will also have high power. Our analysis hypothetically considers a $5\sigma_{\rm n}$ signal~\cite{TASEH2022analysis}. The bins that exceeding $3.355\sigma_{\rm n}$ are determined as signal candidates. In the statistical view, the signal and the noise fluctuation have the probabilities of 0.95 and 0.0004, respectively, to pass this threshold. For a 10-MHz spectrum, the noise fluctuation should contribute 4 candidates on average. The analysis is executed in parallel with the data taking, and candidate bins passing the threshold are singled out~\cite{TASEH2022analysis}. To distinguish the origin of each candidate bin, a rescan procedure is designed and carried out roughly every 10-MHz scan.

To rescan a candidate, $f_{\rm c}$ is tuned to the candidate frequency. The data taking procedure is replicated with the same data integration time $t$ to accumulate more data. The power of the candidate bin is examined afterward. If the power goes above $5\sigma_{\rm n}$, the bin displays a sign of being a signal. If the power goes below $3.355\sigma_{\rm n}$, the candidacy of the bin is removed. If the power is still within $ 3.355\sigma_{\rm n} -  5\sigma_{\rm n}$, the rescan continues until the power goes above $5\sigma_{\rm n}$ or below $3.355\sigma_{\rm n}$. After all the candidates are confirmed, the ordinary scan continues. 

Two further experiments are checked for the confirmed signals to accept their axion related origin. A horn antenna is used to search the external interference at the signal frequencies near the DR. If the signal can not be identified by the horn antenna experiment, a rescan at zero mgnetic field is carried out to see its existence. If the signal remains, its possibility of being the axion signal is ruled out.

\subsection{Synthetic axion experiment}\label{subsec:Synthetic axion experiment}

\begin{figure}[tb]
    \centering
    \includegraphics[width=0.48\textwidth]{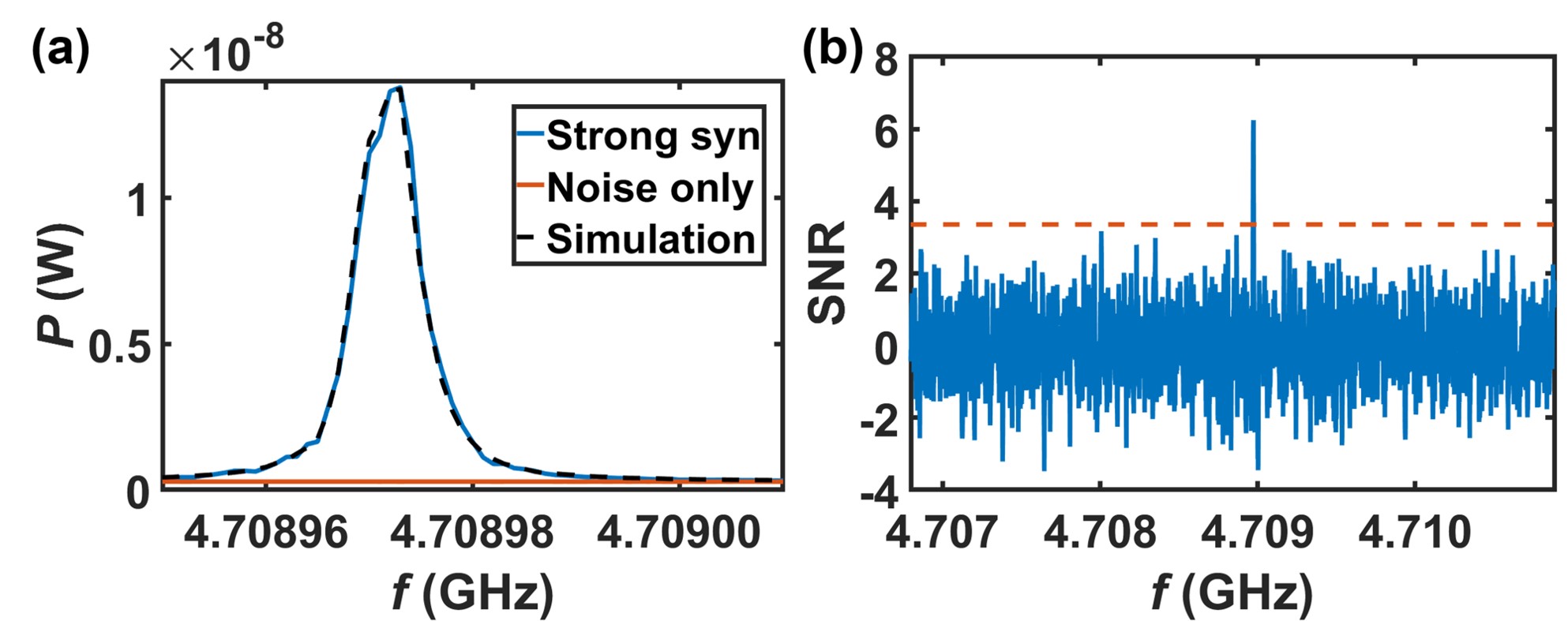}
    \caption{Synthetic axion experiment. (a) Power spectrum of strong synthetic signal with $f_{\rm syn} = 4.708970\ {\rm GHz}$. The blue curve is the data from 3-minute data-taking. The dashed curve is the simulation result of the synthetic signal. The signal from the VST shows a good agreement with the simulation. The red line is the data without the synthetic signal, showing only the system noise. (b) SNR spectrum of weak synthetic axion experiment. An identical synthetic signal as (a) but attenuated by a factor of $2.24 \times 10^4$ is applied to the cavity. A narrow peak with a value of 6.12 at 4.708970 GHz is found.}
    \label{fig:Strong Faxion}
\end{figure}

To test the capability of the experimental setup and the analysis strategy to discover a signal from axion with roughly 10 $g_{\rm a\gamma\gamma}^{\rm KSVZ}$, a synthetic signal is injected to the cavity and detected by our standard axion data taking and analysis procedures~\cite{TASEH2022analysis}. The signal is generated by the VST and applied through the input line 1 to the cavity, and through the output amplification chain 2 back to the VST; see Fig. \ref{fig:MWwiring} for the signal wiring. Figure \ref{fig:Strong Faxion}(a) shows a strong synthetic signal with $f_{\rm syn} = f_{\rm c}$ = 4.708970 GHz read out by the signal receiver. The power of the synthetic signal is assessed based on the calibration of the output probe 2, $T_{\rm sys}$ = 2.4 K and $G_2$ = 99.25 dB = $8.42 \times 10^{9}$. The noise power in a 1-kHz bin read by the VST is $2.79 \times 10^{-10}$ W, corresponding to $P_{\rm n} = 3.31 \times 10^{-20}$ W referred back to the cavity output probe 2. Similarly, the total power of the synthetic signal read by the VST is $1.20 \times 10^{-7}$ W, corresponding to $P_{\rm a} = 1.43 \times 10^{-17}$ W at the cavity output probe 2. With the known $\beta_2$ = 2.35, $P_{\rm syn} = 2.04 \times 10^{-17}$ W is derived, corresponding to an axion signal with 3030 $g_{\rm a\gamma\gamma}^{\rm KSVZ}$. 69.4\% of the total signal power is spread in frequency with a 8 kHz width. The bin with the maximal signal has 11.3\% of the signal power.

In the CD102 run, an identical synthetic signal, but with a factor of 43.50 dB = $2.24 \times 10^4$ power reduction, was sent to test our data analysis. This weak synthetic signal corresponds to an axion signal with $20.2 g_{\rm a\gamma\gamma}^{\rm KSVZ}$. The operation of the synthetic signal data taking was exactly the same as the normal axion scan. With the integration time $t$ = 40 minute, the anticipated SNR of the maximal power bin is 3.36. Via a Monte Carlo simulation, the SNR is expected to increase to 6.9 with a variation of 0.8 after applying our standard analysis procedures~\cite{TASEH2022analysis}. Figure \ref{fig:Strong Faxion}(b) displays the analysis result of the weak synthetic signal experiment. The SNR spectrum has a narrow peak at 4.708970 GHz with a value of 6.12~\cite{TASEH2022analysis}, indeed compatible to the anticipated value. The simulation can reliably reproduce the weak synthetic signal experiment result from the physical data taking instruments and the data analysis. Similar simulations demonstrate that the system is capable of detecting an approximately 10$g_{\rm a\gamma\gamma}^{\rm KSVZ}$ axion signal assuming that 95\% of the signal power concentrates in 5 frequency bins.

\subsection{Unexpected experiment issues}

During the data taking operation, a magnet quench occurred at 9:20am on October 16, 2021, due to a cooling water failure. No damage arose in the experimental setup. The cooling water issue was resolved after 3 days. The DR and the magnet returned to the normal operation after another 3 days, and after that the data taking resumed within 3 hours without difficulty. 

An earthquake of the intensity scale 4 stroke the lab at 1:11pm on October 24. After the earthquake the DR temperature $T_{\rm mx}$ increased to 38 mK and then dropped back to 27 mK. The data taking was not affected. 

These accidents, including the magnetic quench and the earthquake, induced changes of the data taking readout power $P_2$ by a maximum of 4\% for $\approx$ 10\% of the data taking time. At this moment the source of the changes is not understood, and therefore the probable corresponding changes of $T_{\rm sys}$ is quoted as a systematic uncertainty in the analysis~\cite{TASEH2022analysis}.
\section{Result}

\begin{figure*}[tb]
    \centering
    \includegraphics[width=0.7\textwidth]{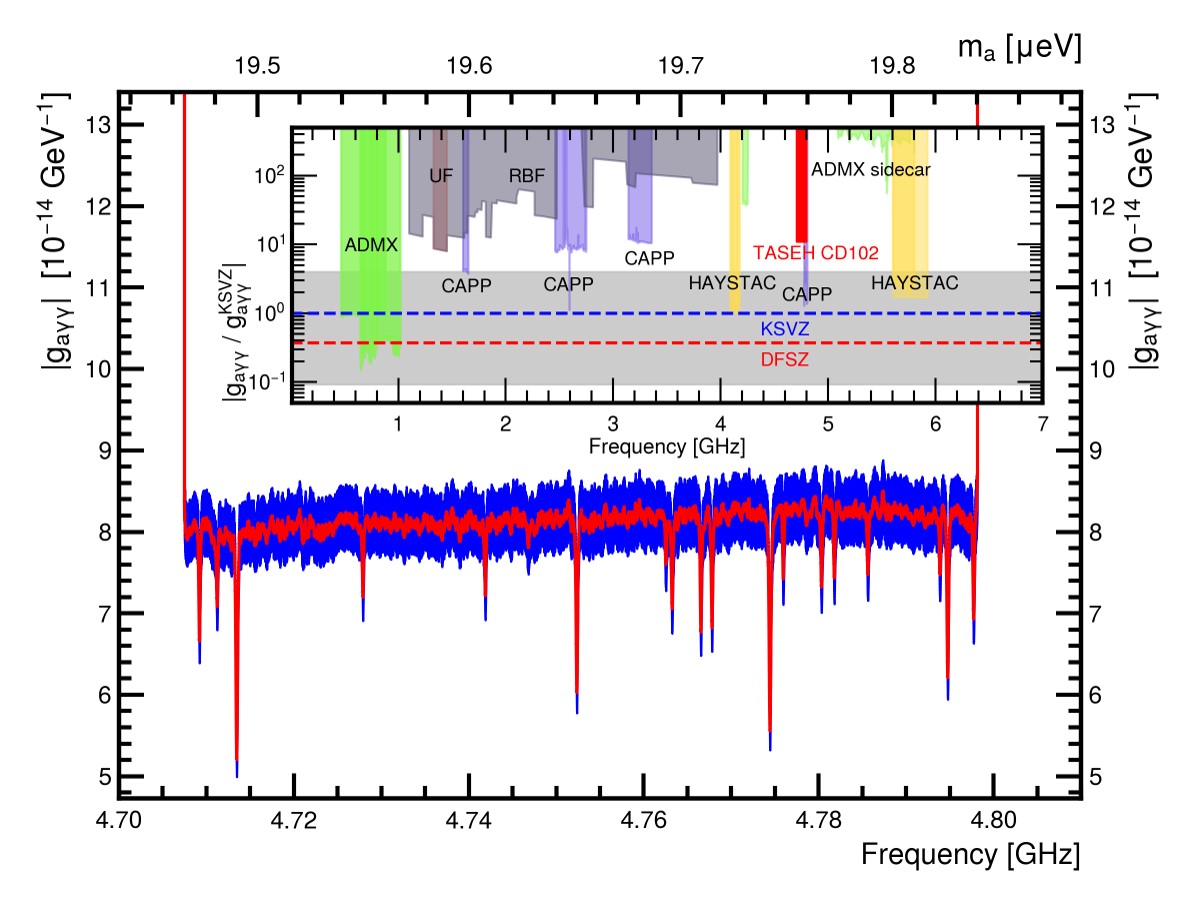}
    \caption{Exclusion plot of $g_{\rm a\gamma\gamma}$ from TASEH CD102 data. The limits on $|g_{\rm a\gamma\gamma}|$ for the frequency range of $4.70750 - 4.79815$ GHz. Inset compares the current result with previous searches performed by the ADMX, CAPP, HAYSTAC, RBF (Rochester-Brookhaven-Fermilab), and UF (University of Florida) Collaborations. The gray band shows the allowed region of $|g_{\rm a\gamma\gamma}|$ vs. $m_{\rm a}$ from various QCD axion models, while the blue and red dashed lines are the values predicted by the KSVZ and DFSZ benchmark models, respectively.}
    \label{fig:axionlimit}
\end{figure*}

The described system was cooled down in mid-October 2021 (the CD102 run) for the axion search. The search operation started on October 13. The first step was at $\theta$ = 99.4\textdegree, corresponding to $f_{\rm c}$ = 4.79815 GHz. The angle was running backward for 837 steps to scan $f_{\rm c}$ over the $4.70750-4.79815$ GHz range. The last step ended at $\theta$ = 58.5\textdegree~on November 15. The data took 55 TB storage space.

In the scan range, 22 bins had SNR greater than 3.355. 20 candidates were ruled out since their SNR dropped below 3.355 after the rescan procedure. The remaining 2 candidates in the frequency ranges of $4.71017-4.71019$ and $4.74730-4.74738$ GHz had SNR larger than 5 after the rescan. However, none of them was considered as an axion signal. The signals in the second frequency range were detected outside the DR by the horn antenna. They were from the instrument control computer. The signals in the first range were weaker, and no external interference signal was found. However, the signals were still alive at zero magnetic field. Hence, no axion signal was found in the $4.70750-4.79815$ GHz range, corresponding to the $19.4687-19.8436\ \si{\micro \eV}$ mass range. Figure \ref{fig:axionlimit} shows the axion-photon coupling constant $g_{\rm a\gamma\gamma}$ exclusion plot in this range. The result excluded $g_{\rm a\gamma\gamma} \gtrsim 8.1 \times 10^{-14} \ {\rm GeV}^{-1}$, corresponding to a factor 11 of the KSVZ banchmark model, at the 95\% confidence level. Note that no limit was placed in the $4.71017-4.71019$ GHz and $4.74730-4.74738$ GHz ranges as the external interference existed during the data taking. The details of the analysis can be found in our parallel analysis paper~\cite{TASEH2022analysis}.
\section{Conclusion and future plan}

The Taiwan Axion Search Experiment with Haloscope (TASEH) collaboration has built an axion search haloscope. The system includes a dilution refrigerator, hosting a frequency tunable cavity detector with volume of 0.234 liter in a magnet of 8-T nominal field, and a signal receiver of system noise temperature of 2.2 K. The system is working and fully calibrated. The TASEH Collaboration performed the first axion search experiment in the mass range of $19.4687-19.8436\ \si{\micro \eV}$. The search excludes values of the axion-photon coupling constant $g_{\rm a\gamma\gamma}$ 11 times above the KSVZ benchmark model at the 95\% confidence level. The result concludes the first phase of the TASEH haloscope development.

The next phase of the development will be integrating a quantum-limited Josephson parametric amplifier to the readout amplification chain. A factor of $\sim$8 improvement in the detection signal-to-noise ratio (SNR) is expected from this integration. A magnet of 9 T field and 152 mm bore has been procured. Possible cavity designs with larger volume are being investigated with a goal to improve the SNR by a factor of $\sim$8. The combined improvements will enable the TASEH collaboration to contribute the axion search efforts in the $10-25\ \si{\micro \eV}$ mass range to the QCD axion-photon coupling limit.

\begin{acknowledgments}

The authors are from the TASEH collaboration. We thank Chao-Lin Kuo for his help to initiate this project as well as discussions on the microwave cavity design, Gray Rybka and Nicole Crisosto for their introduction of the ADMX experimental setup and analysis, Anson Hook for the discussions and the review of the axion theory, and Jiunn-Wei Chen, Cheng-Wei Chiang, Cheng-Pang Liu, and Asuka Ito for the discussions of future improvements in axion searches. The authors acknowledge the support of microwave test and measurement equipment from the National Chung-Shan Institute of Science and Technology, the dilution refrigerator from the Instrument Center, National Chung Hsing University, and the computational and storage resources from the National Center for High-performance Computing of National Applied Research Laboratories in Taiwan. The work of the TASEH Collaboration was funded by the Ministry of Science and Technology (MoST) of Taiwan with grant numbers MoST-109-2123-M-001-002, MoST-110-2123-M-001-006, MoST-110-2112-M-213-018, MoST-110-2628-M-008-003-MY3, and MoST-109-2112-M-008-013-MY3, and by the Institute of Physics, Academia Sinica. 

\end{acknowledgments}

\appendix
\section{Tables of experimental setup}\label{sec:Table}

The tables for the DR thermometer and dc wiring details and the MW system component part numbers are listed.

\begin{table}[ht]
    \centering
    \begin{tabular}{|*{5}{c|}}
    \hline
        \multirow{2}{*}{\centering Channel} & \multirow{2}{*}{\centering Location} & \multirow{2}{*}{\centering Type} & \multirow{2}{1.5cm}{\centering Cal. range $\rm [Kelvin]$} & \multirow{2}{1.5cm}{\centering Excitation level $[\si{\micro\volt}]$} \\&&&&\\ \hline
        1 & 50K flange & Platinum & 310-20 & 2000 \\ \hline
        2 & 4K flange & Cernox & 310-0.1 & 632 \\ \hline
        3 & Magnet & Cernox & 310-0.1 & 632 \\ \hline
        5 & Still flange & Cernox & 310-0.1 & 200 \\ \hline
        6 & Mixing flange & $\rm RuO_2$ & 100-0.007 & 20 \\ \hline
        7 & Cavity & $\rm RuO_2$ & 90-0.03 & 20 \\ \hline
        8 & Calibration plate & Cernox & 90-0.06 & 632 \\ \hline
    \end{tabular}
    \caption{DR thermometer detailed information.}
    \label{table:tablethermometer}
\end{table}

\begin{table}[h]
    \centering
    \begin{tabular}{|*{4}{c|}}
    \hline
        Channel & RT-4K wiring & 4K-MX wiring & Resistance $\rm [\Omega]$ \\ \hline
        1-8 & 42 copper & 42 copper & 0.25 \\ \hline
        9-24 & 38 constantan & 45 manganin & 180 \\ \hline
        25-48 & 38 constantan & 42 NbTi & 30 \\ \hline
    \end{tabular}
    \caption{dc wiring material and resistance in DR. The numbers (Standard Wire Gauge) next to the materials denote the wire sizes. The resistance values are in the cryogenic operation. Each channel consists of a RT-4K wire and a 4K-MX wire in series.}
    \label{table:tabledcwiring}
\end{table}

\begin{table}[h]
    \centering
    \begin{tabular}{|l|l|l|}
    \hline
         Label & Description & \textbf{Supplier} Model/Series No. \\ \hline
         A1 & HEMT amplifier & \textbf{CIT} cryo4-12 \\ \hline
         A2 & HEMT amplifier & \textbf{LNF} LNF-LNC4\_8C \\ \hline
         A3 & Low-noise amplifier & \textbf{Mini-Circuits} ZX60-83LN-S+ \\ \hline
         At1 & 10dB cryo attenuator & \textbf{XMA} 2082-6043-10-cryo \\ \hline
         At2 & 3dB cryo attenuator & \textbf{XMA} 2082-6043-3-cryo \\ \hline
         At3 & 20dB cryo attenuator & \textbf{XMA} 2082-6043-20-cryo \\ \hline
         BS & 50$\Omega$ terminator & \textbf{XMA} 2003-6117-00-cryo \\ \hline
         C & Circulator & \textbf{LNF} CIISISC4\_8A \\ \hline
         Co1 & Semi-flexible Cu coax & \textbf{Johnson} 415-0081 \\ \hline
         Co2 & Semi-rigid Cu coax & \textbf{Woken} 00100A1G2A197C \\ \hline
         Co3 & Semi-rigid NbTi coax & \textbf{Coax} SC-219/50-NbTi-NbTi \\ \hline
         \multirow{2}{*}{Co4} & \multirow{2}{3cm}{Semi-rigid Ag-plated CuNi coax} & \multirow{2}{*}{\textbf{Coax} SC-219/50-SCN-CN} \\&&\\ \hline
         S1 & SPDT switch & \textbf{Radiall} R570463000 \\ \hline
         S2 & SPDT relay module & \textbf{NI} PXIe-2599 \\ \hline
         VST & Vec. signal transceiver & \textbf{NI} PXIe-5644R \\ \hline
         VNA & Vec. network analyzer & \textbf{HP} 8720B \\ \hline
    \end{tabular}
    \caption{MW instrument models and component part numbers.}
    \label{tab:component}
\end{table}

\phantomsection
\addcontentsline{toc}{chapter}{Bibliography}
\nocite{*}
\bibliographystyle{apsrev}
\bibliography{mybib.bib}

\end{document}